\documentclass[aps,preprint,showpacs,groupedaddress,floatfix]{revtex4}
\usepackage{graphicx}
\usepackage{dcolumn}
\usepackage{bm}
\begin{document}
\title{Relativistic Hartree-Bogoliubov model with
density-dependent meson-nucleon couplings}
\author{T. Nik\v si\' c}
\author{D. Vretenar}
\affiliation{ Physics Department, Faculty of Science, University of
Zagreb, 10 000 Zagreb, Croatia}

\author{P. Finelli}
\affiliation{Physics Department, University of Bologna, and INFN - Bologna,
I-40126 Bologna, Italy}

\author{P. Ring}
\affiliation{Physik-Department der Technischen Universit\"at M\"unchen,
D-85748 Garching, Germany}
\vspace{1cm}
\date{\today}

\begin{abstract}
The relativistic Hartree-Bogoliubov (RHB) model is extended to include
density dependent meson-nucleon couplings. The effective Lagrangian is
characterized by a phenomenological density dependence for the $\sigma$,
$\omega$ and $\rho$ meson-nucleon vertex functions, adjusted to properties
of nuclear matter and finite nuclei. Pairing correlations are described by
the pairing part of the finite range Gogny interaction. The new
density-dependent effective interaction DD-ME1
is tested in the analysis of the equations of state for symmetric
and asymmetric nuclear matter, and of ground-state properties of the
Sn and Pb isotopic chains. Results of self-consistent RHB calculations
are compared with experimental data, and with results previously
obtained in the RHB model with non-linear self-interactions, as well as
in the density dependent relativistic hadron field (DDRH) model.
Parity-violating elastic electron scattering on Pb and Sn nuclei
is calculated using a relativistic optical model with inclusion of
Coulomb distortion effects, and the resulting asymmetry parameters
are related to the neutron ground-state density distributions.
\end{abstract}
\pacs{21.60.-n, 21.30.Fe, 21.65.+f, 21.10.-k}

\maketitle

\section{\label{secI}Introduction}
Models based on concepts of nonrenormalizable
effective relativistic field theories and density functional theory
provide a rich theoretical framework for studies of
nuclear structure phenomena, not only in nuclei along the valley of
$\beta$-stability, but also in exotic nuclei with extreme isospin
values and close to the particle drip lines.
A well known example of an effective theory of nuclear structure is
Quantum Hadrodynamics (QHD), a field theoretical
framework of Lorentz-covariant, meson-nucleon or point-coupling
models of nuclear dynamics~\cite{SW.97}.
The effective Lagrangians of QHD consist of known long-range interactions
constrained by symmetries
and a complete set of generic short-range interactions.
The QHD framework implicitly includes vacuum effects, chiral symmetry,
nucleon substructure, exchange terms, long- and short-range correlation
effects.

Structure models based on the relativistic mean-field (RMF)
approximation have been successfully employed in studies
of spherical and deformed nuclei
all over the periodic table~\cite{Rin.96}. The relativistic
framework has also been applied in studies of
nuclear structure phenomena in nuclei far from $\beta$- stability.
In particular, in studies of isotopic chains that also included
exotic nuclei with extreme isospin values,
we have used the relativistic Hartree-Bogoliubov (RHB) model,
which encloses a unified description of mean-field and pairing
correlations. A number of interesting nuclear properties have been
studied with the RHB model:
the halo phenomenon in light nuclei~\cite{PVL.97},
properties of light neutron-rich nuclei~\cite{LVP.98},
the reduction of the effective single-nucleon spin-orbit
potential in nuclei close to the drip-lines~\cite{LVR.97},
properties of neutron-rich Ni and Sn isotopes~\cite{LVR.98},
the location of the proton drip-line between
from $Z=31$ to $Z=73$ and the phenomenon of
ground-state proton radioactivity ~\cite{VLR.99,LVR.99,LVR.01b}.

The details of calculated nuclear properties that can be
compared with empirical data, as well as the predictions of new
phenomena far from $\beta$-stability (halo nuclei, neutron skins,
suppression of shell effects in neutron rich nuclei,
ground-state proton radioactivity
beyond the drip-line, the onset of exotic collective modes),
crucially depend on the choice of the effective RMF Lagrangian
in the $ph$ channel, as well as on the treatment of pairing
correlations. Several phenomenological parameterizations
of the effective Lagrangian have been derived that
provide a satisfactory description of nuclear properties along
the $\beta$-stability line, as well as in regions far from
stability. These effective interactions are characterized by a
minimal set of model parameters: meson masses and meson-nucleon
coupling constants. The most successful RMF effective interactions
are purely phenomenological, with parameters adjusted to reproduce
the nuclear matter equation of state and a set of global
properties of spherical closed-shell nuclei.
In most applications of the RHB model, in particular, we have
used  the NL3 effective interaction \cite{LKR.97} for the RMF effective
Lagrangian. Properties calculated with NL3 indicate that this is probably
the best effective interaction so far, both for nuclei at and away from the
line of $\beta $-stability.

The limitations of standard RMF effective interactions are, however, well
known, even for nuclei close to the stability line. They are more
pronounced in the isovector channel, which is poorly constrained
by the available experimental data on ground-state properties of
nuclei. For example, in a recent analysis of neutron
radii in the framework of mean-field models~\cite{Fur.01}, it has been shown
that conventional RMF models systematically overestimate the values
of $r_n - r_p$. It is well known that they also predict an equation of
state of neutron matter that is very different from the standard
microscopic many-body neutron matter equation of state of
Friedman and Pandharipande~\cite{FP.81}. The parameterization of the
isovector channel of an effective RMF interaction is, of course,
extremely important for possible extrapolations to neutron-rich or
proton-rich nuclei. For example, various
non-relativistic and relativistic mean-field models
differ significantly in the prediction of the two-neutron separation
energies of the Sn isotopes with $A > 132$~\cite{Naz.99} and, therefore,
of the exact location of the drip-line. There are also differences in the
predicted location of the drip-line on the proton-rich side~\cite{LVR.01b}.
The properties of effective interactions are very important
for the description of phenomena in nuclear astrophysics. The choice
of the effective RMF Lagrangian directly determines the calculated
properties of neutron stars (radii, surface crust)~\cite{HP.01,HP.02},
as well as the properties of nuclei that take part in the r-process
or rp-process of nucleosynthesis.

In order to overcome the limitations of standard RMF models,
several solutions have been put forward. An obvious choice is to
extend the minimal six-parameter RMF model by including additional
interaction terms in the isoscalar, as well as in the isovector
channel. The complex many-body dynamics is effectively included in
additional non-linear self-interactions. This approach has been
studied both in the framework of meson-exchange
models~\cite{Bod.91,ST.94,SFM.00,HP.02} and with relativistic
point-coupling models~\cite{RF.97,BMM.02}. Even though some
interesting results have been obtained, especially in applications
to nuclear astrophysics, the situation is not satisfactory. The
problem is that the empirical data set of bulk and single-particle
properties of finite nuclei can only constrain six or seven
parameters in the general expansion of an effective
Lagrangian~\cite{FS.00}. One can, of course, include interaction
terms that describe specific phenomena, but their coupling
parameters and even their forms cannot be accurately determined in
this way. In some cases even the signs of interaction terms are
not determined by the data. The general expansion of an effective
Lagrangian in powers of the fields and their derivatives can be
controlled by the ``naive dimensional analysis"
(NDA)~\cite{FML.96,FST.97,RF.97,BMM.02}. NDA tests the
coefficients of the expansion for ``naturalness", i.e. this method
controls the magnitude of the coupling constants. Although NDA can
exclude some interaction terms because their couplings would be
``unnatural", it cannot determine the parameters of a model on a
level of accuracy that is required for a quantitative analysis of
nuclear structure data.

Instead of including additional non-linear self-interaction terms
in effective RMF Lagrangians, another possibility is to formulate
an effective hadron field theory with medium dependent
meson-nucleon vertices. Such an approach retains the basic
structure of QHD, but could be more directly related to the
underlying microscopic description of nuclear interactions. In the
density dependent relativistic hadron field (DDRH) model of
Refs.~\cite{BT.92,FL.95,FLW.95} the medium dependence of the
vertices is expressed by a functional of the baryon field
operators. The meson-baryon coupling constants in nuclear matter
are adjusted to the Dirac-Brueckner (DB) self-energies. A
Lorentz-invariant functional is defined to project the nuclear
matter results onto the meson-nucleon vertices of the effective
DDRH model for finite nuclei. In the early version~\cite{BT.92}
of this model, the density dependence was only included in the field
equations after the variation. 
However, a consistent treatment of medium effects
implies a variation of the vertex functionals with respect to the
baryon field operators, and this results in additional
rearrangement self-energies in the single-nucleon Dirac equation
~\cite{FLW.95}. In Ref.~\cite{FLW.95} the model was applied in the
calculation of ground-state properties of doubly-closed shell
nuclei. Density dependent $\sigma$ and $\omega$ meson couplings
were used that were derived from DB calculations using the Bonn
A, B, and C nucleon-nucleon potentials. It was shown that the
inclusion of rearrangement self-energies is essential for a
quantitative description of bulk properties and single-particle
spectra. The model has been recently extended to
hypernuclei~\cite{KHL.00}, neutron star matter~\cite{HKL.01a}, and
asymmetric nuclear matter and exotic nuclei~\cite{HKL.01}. The
density dependent interactions have been derived from the
Groningen and Bonn-A nucleon-nucleon potentials. For finite
nuclei, in particular, the quality of the calculated properties
are comparable with non-linear RMF models.

In Ref.~\cite{TW.99} Typel and Wolter introduced a phenomenological
density dependence for the $\sigma$, $\omega$ and $\rho$ meson-nucleon
couplings, adjusted to properties of nuclear matter and some finite nuclei.
The parameters of their DDRH model were also compared with coupling
constants derived from DB calculations of nucleon self-energies. The model
was used to study the equation of state of symmetric and asymmetric nuclear
matter, and the ground state properties of semi-closed shell nuclei.
Even though pairing correlations were only treated in the BCS approximation,
properties at the proton and neutron drip-lines were calculated. It was,
however, emphasized that a more realistic description of nuclei
at the drip-lines necessitates the use of the Hartree-Bogoliubov
framework with finite range pairing interactions. The phenomenological
ansatz of Ref.~\cite{TW.99} for the functional form of the density
dependence of the meson-nucleon vertices was also used in
Ref.~\cite{HKL.01} to derive the meson-nucleon coupling parameters
from the Groningen and Bonn-A NN potentials.

In this work we present an extension of the relativistic Hartree-Bogoliubov
(RHB) model that includes density dependent meson-nucleon couplings.
For the effective RMF Lagrangian we follow the approach of
Typel and Wolter~\cite{TW.99} and use their phenomenological
functional forms for the density dependence of the vertex functions.
The parameters of the effective interaction are, however, adjusted
in a different way. Pairing correlations are described by the
pairing part of the finite range Gogny interaction. The new model
is tested in the analysis of the equation of state for symmetric
and asymmetric nuclear matter, and ground-state properties of the
Sn and Pb isotopic chains. The results are compared with experimental
data and with previous results obtained in the RHB, as well as
DDRH frameworks.

In Sec.~\ref{secII} we outline the RHB model with
density dependent meson-nucleon couplings. The new parameterization
of the vector density dependence of the vertex functionals
is discussed in Sec.~\ref{secIII} in comparison with the
effective interaction of Typel and Wolter, and we also analyze
the results for the equation of state for symmetric and asymmetric
nuclear matter. In Sec.~\ref{secIV} the new density-dependent
effective interaction DD-ME1 is employed in the RHB calculations of
ground-state properties of Sn and Pb nuclei. Binding energies,
charge radii, differences between neutron and proton radii,
spin-orbit splittings, and charge isotope shifts are compared
with available experimental data, and with the results obtained
with the non-linear interaction NL3. Sec.~\ref{secV} contains an
analysis of parity-violating elastic electron scattering on
$^{208}$Pb and on those Sn isotopes for which there are experimental
data on $r_n - r_p$. For the elastic scattering of 850 MeV electrons
on these nuclei, the calculated parity-violating asymmetry parameters
are related to the Fourier transforms of the neutron density
distributions. In Sec.~\ref{secVI} we summarize the results of this
work and present an outlook for future applications of the
RHB model with density-dependent meson-nucleon couplings.


\section{\label{secII}Relativistic Hartree-Bogoliubov model with
density-dependent meson-nucleon couplings}
The framework of density-dependent hadron field theory is described in
great detail in Refs.~\cite{FLW.95,TW.99,HKL.01}. In this section we
outline the essential features of the model with vector density dependence
of the meson-nucleon couplings. The model is defined by the relativistic
Lagrangian density
\begin{eqnarray}
{\cal L} &=&\bar{\psi}\left( i{\bm \gamma} \cdot 
{\bm \partial} -m\right) \psi ~+~\frac{1%
}{2}(\partial \sigma )^{2}-\frac{1}{2}m_{\sigma }\sigma ^{2}  \nonumber \\
&&-~\frac{1}{4}\Omega _{\mu \nu }\Omega ^{\mu \nu }+\frac{1}{2}m_{\omega
}^{2}\omega ^{2}~-~\frac{1}{4}{\vec{{\rm R}}}_{\mu \nu }{\vec{{\rm R}}}^{\mu
\nu }+\frac{1}{2}m_{\rho }^{2}\vec{\rho}^{\,2}~-~\frac{1}{4}{\rm F}_{\mu \nu
}{\rm F}^{\mu \nu }  \nonumber \\
&&-~g_{\sigma }\bar{\psi}\sigma \psi ~-~g_{\omega }\bar{\psi}\gamma \cdot
\omega \psi ~-~g_{\rho }\bar{\psi}\gamma \cdot \vec{\rho}\vec{\tau}\psi ~-~e%
\bar{\psi}\gamma \cdot A\frac{(1-\tau _{3})}{2}\psi \;.
\label{lagrangian}
\end{eqnarray}
Vectors in isospin space are denoted by arrows, and bold-faced
symbols will indicate vectors in ordinary three-dimensional space.
The Dirac spinor $\psi$ denotes
the nucleon with mass $m$.  $m_\sigma$, $m_\omega$, and
$m_\rho$ are the masses of the $\sigma$-meson, the
$\omega$-meson, and the $\rho$-meson.  $g_\sigma$,
$g_\omega$, and $g_\rho$ are the corresponding coupling
constants for the mesons to the nucleon. $e^2 /4 \pi =
1/137.036$.  The coupling constants and unknown meson masses
are parameters, adjusted to reproduce nuclear matter properties and
ground-state properties of finite nuclei.
$\Omega ^{\mu \nu }$, $\vec{R}^{\mu \nu }$, and $F^{\mu \nu }$ are the field
tensors of the vector fields $\omega $, $\rho $, and of the photon:
\begin{eqnarray}
\Omega ^{\mu \nu } &=&\partial ^{\mu }\omega ^{\nu }-\partial ^{\nu }\omega
^{\mu } \\
\vec{R}^{\mu \nu } &=&\partial ^{\mu }\vec{\rho}^{\,\nu }-\partial ^{\nu }%
\vec{\rho}^{\,\mu } \\
F^{\mu \nu } &=&\partial ^{\mu }A^{\nu }-\partial ^{\nu }A^{\mu } \;.
\end{eqnarray}
$g_\sigma$, $g_\omega$, and $g_\rho$ are assumed to be vertex functions
of Lorentz-scalar bilinear forms of the nucleon operators. In most
applications of the density-dependent hadron field theory the meson-nucleon
couplings are functions of the vector density
\begin{equation}
\rho _v = \sqrt{j_{\mu} j^{\mu}},~~~~~~{\rm with}~~~~
j_{\mu} = \bar{\psi} \gamma _{\mu} \psi \;.
\end{equation}
Another obvious choice is the dependence on the scalar density
$\rho_s = \bar{\psi} \psi$. It has been shown, however, that
the vector density dependence produces better results for
finite nuclei~\cite{FLW.95}, and provides a more natural
relation between the self-energies of the density-dependent
hadron field theory and the Dirac-Brueckner microscopic
self-energies~\cite{HKL.01}. In the following we assume
the vector density dependence of the meson-nucleon couplings.
The single-nucleon Dirac equation is derived by variation of the
Lagrangian (\ref{lagrangian}) with respect to $\bar{\psi}$
\begin{equation}
\left [ \gamma^{\mu} (i \partial _{\mu} - \Sigma  _{\mu})
- (m - \Sigma) \right ] \psi = 0 \;,
\end{equation}
with the nucleon self-energies defined by the following relations
\begin{equation}
\Sigma = g_{\sigma } \sigma
\end{equation}
\begin{equation}
\Sigma  _{\mu} = g_{\omega } \omega _{\mu} +
g_{\rho } \vec{\tau}\cdot \vec{\rho} _{\mu} +
e \frac{(1-\tau _{3})}{2} A _{\mu} + \Sigma^R  _{\mu} \;.
\end{equation}
The density dependence of the vertex functions
$g_\sigma$, $g_\omega$, and $g_\rho$ produces the
{\it rearrangement} contribution $\Sigma_R^{\mu}$ to
the vector self-energy
\begin{equation}
\Sigma^R  _{\mu} =
 \frac{j_{\mu}}{\rho _v} \left(
  \frac {\partial g_{\omega }}{\partial \rho _v} \bar{\psi} \gamma ^{\nu }
  \psi \omega _{\nu} +
  \frac {\partial g_{\rho }}{\partial \rho _v} \bar{\psi} \gamma ^{\nu }
  \vec{\tau} \psi \cdot \vec{\rho} _{\nu} +
  \frac {\partial g_{\sigma }}{\partial \rho _v} \bar{\psi} \psi \sigma
\right ).
\end{equation}
The inclusion of the rearrangement self-energies
is essential for the energy-momentum conservation and the
thermodynamical consistency of the model~\cite{FLW.95,TW.99}.

The lowest order of the quantum field theory is the {\it
mean-field} approximation: the meson field operators are
replaced by their expectation values.  The $A$ nucleons,
described by a Slater determinant $|\Phi\rangle$ of
single-particle spinors $\psi_i,~(i=1,2,...,A)$, move
independently in the classical meson fields.  The sources
of the meson fields are defined by the nucleon densities
and currents. The ground state of a nucleus is described
by the stationary self-consistent solution of the coupled
system of Dirac and Klein-Gordon equations. Due to
charge conservation, only the 3-component of the
isovector rho meson contributes. For an even-even system
the spatial vector components \mbox{\boldmath $\omega,~\rho_3$} and
${\bf A}$ vanish, and the self-energies are determined by the
solutions of the Klein-Gordon and Poisson equations
\begin{eqnarray}
(-\Delta +m_{\sigma })\sigma ({\bf r}) &=&-g_{\sigma }\rho _{s}({\bf r}),
\label{kleingordons} \\
(-\Delta +m_{\omega })\omega ({\bf r}) &=&g_{\omega }\rho _{v} ({\bf r}),
\label{kleingordono} \\
(-\Delta +m_{\rho })\rho _{3}({\bf r}) &=&g_{\rho }(\rho _{n}({\bf r)}-\rho
_{p}({\bf r)}),
\label{kleingordonr} \\
-\Delta A_{0}({\bf r}) &=&e^{2}\rho _{c}({\bf r)} \;.
\label{poisson}
\end{eqnarray}
The source terms on the left-hand side of these equations
are sums of bilinear products of baryon amplitudes.
The densities are calculated in the {\it no-sea} approximation,
i.e. only occupied single-nucleon states with positive energy
explicitly contribute to the nucleon self-energies.

In addition to the self-consistent mean-field
potential, pairing correlations have to be included in order to
describe ground-state properties of open-shell nuclei.
For spherical and deformed nuclei not too far from
the stability line, pairing is
often treated phenomenologically
in the simple BCS approximation \cite{Rin.96}.
However, the BCS model presents only a poor approximation for
exotic nuclei far from the valley of $\beta$-stability.
The structure of weakly bound nuclei
necessitates a unified and self-consistent treatment of
mean-field and pairing correlations. In particular, the
relativistic Hartree-Bogoliubov (RHB)
model~\cite{PVL.97,LVP.98,LVR.98,LVR.99} represents a relativistic
extension of the Hartree-Fock-Bogoliubov (HFB) framework.
In the RHB model the ground state of a nucleus $\vert \Phi >$ is represented
by the product of independent single-quasiparticle states.
These states are eigenvectors of the
generalized single-nucleon Hamiltonian that
contains two average potentials: the self-consistent mean-field
$\hat\Gamma$, which encloses all the long range particle-hole ({\it ph})
correlations, and a pairing field $\hat\Delta$, which sums
up the particle-particle ({\it pp}) correlations.
In the Hartree approximation for
the self-consistent mean field, the relativistic
Hartree-Bogoliubov equations read
\begin{equation}
\left(
\begin{array}{cc}
\hat{h}_{D}-m-\lambda  & \hat{\Delta} \\
-\hat{\Delta}^{\ast } & -\hat{h}_{D}+m+\lambda
\end{array}
\right) \left(
\begin{array}{c}
U_{k}({\bf r}) \\
V_{k}({\bf r})
\end{array}
\right) =E_{k}\left(
\begin{array}{c}
U_{k}({\bf r}) \\
V_{k}({\bf r})
\end{array}
\right)
\label{HFB}
\end{equation}
where $\hat{h}_{D}$ is the single-nucleon Dirac Hamiltonian,
and $m$ is the nucleon mass. The chemical potential $\lambda $ has to be
determined by the particle number subsidiary condition, in order that the
expectation value of the particle number operator in the ground state equals
the number of nucleons. The column vectors denote the quasiparticle wave
functions, and $E_{k}$ are the quasiparticle energies.
The source terms in equations (\ref{kleingordons}) to (\ref
{poisson}) are sums of bilinear products of baryon amplitudes
\begin{eqnarray}
\rho _{s}({\bf r}) &=&\sum\limits_{E_{k}>0}V_{k}^{\dagger }({\bf r})\gamma
^{0}V_{k}({\bf r}),  \label{equ.2.3.h} \\
\rho _{v}({\bf r}) &=&\sum\limits_{E_{k}>0}V_{k}^{\dagger }({\bf r})V_{k}(%
{\bf r}), \\
\rho _{n}({\bf r)}-\rho_{p}({\bf r)}&=&
\sum\limits_{E_{k}>0}V_{k}^{\dagger }({\bf r})\tau_{3}V_{k}({\bf r}), \\
\rho _{{c}}({\bf r}) &=&\sum\limits_{E_{k}>0}V_{k}^{\dagger }({\bf r}){%
\frac{{1-\tau _{3}}}{2}}V_{k}({\bf r}),
\end{eqnarray}
where the sums run over all positive energy states.
In most applications of the RHB model a phenomenological pairing
interaction has been used, the pairing part of the Gogny force,
\begin{equation}
V^{pp}(1,2)~=~\sum_{i=1,2}e^{-(({\bf r}_{1}-{\bf r}_{2})/{\mu _{i}}%
)^{2}}\,(W_{i}~+~B_{i}P^{\sigma }-H_{i}P^{\tau }-M_{i}P^{\sigma }P^{\tau }),
\end{equation}
with the set D1S \cite{BGG.84} for the parameters $\mu _{i}$, $W_{i}$,
$B_{i}$, $H_{i}$, and $M_{i}$ $(i=1,2)$.

The RHB equations are solved self-consistently, with
potentials determined in the mean-field approximation from
solutions of Klein-Gordon equations for the meson fields.
The Dirac-Hartree-Bogoliubov equations and the equations for the
meson fields are solved by expanding the nucleon spinors
$U_k({\bf r})$ and $V_k({\bf r})$,
and the meson fields in terms of the eigenfunctions of a
spherical or deformed axially symmetric oscillator potential.
A detailed description of the relativistic Hartree-Bogoliubov
model for spherical and deformed nuclei can be found in
Refs.~\cite{LVP.98} and \cite{LVR.99}, respectively.

\section{\label{secIII}Parameterization of the density dependence
of the meson-nucleon couplings}
The density dependence of the meson-nucleon couplings can be obtained
from microscopic Dirac-Brueckner (DB) calculations of nucleon-self energies
in symmetric and asymmetric nuclear matter~\cite{FLW.95,JL.98}.
However, depending
on the choice of the nucleon-nucleon potential and on the approximations
in the DB calculation, rather different results are obtained
for the density dependence of the vertex functions. At low densities,
in particular, DB calculations of nuclear matter become unreliable, and
meson-nucleon couplings determined directly from DB self-energies provide
only a qualitative description of ground-state properties for finite nuclei.
Instead of adjusting the vertex functions directly to DB self-energies,
in Ref.~\cite{TW.99} an ansatz was made for the functional form of
the density dependence that encloses different DB results. The parameters
of the density dependence were obtained from a fit to properties of
nuclear matter and finite nuclei. The same functional form was also
used in Ref.~\cite{HKL.01} to adjust the meson-nucleon couplings to the
DB self-energies derived from the Groningen and Bonn-A nucleon-nucleon
potentials. In this work we adopt for the density dependence of the
meson-nucleon couplings the functionals of Ref.~\cite{TW.99}, and adjust
the parameters to properties of symmetric and asymmetric nuclear matter,
binding energies, charge radii and neutron radii of spherical nuclei.
The coupling of the $\sigma$-meson and $\omega$-meson to the nucleon
field reads
\begin{equation}
g_i(\rho) = g_i(\rho_{\rm sat}) f_i(x)\quad {\rm for}\quad i=\sigma, \omega\;,
\end{equation}
where
\begin{equation}
f_i(x) = a_i \frac{1 + b_i(x+d_i)^2}{1 + c_i(x+d_i)^2}
\label{func}
\end{equation}
is a function of $x = \rho /\rho_{\rm sat}$. The eight real parameters
in (\ref{func}) are not independent. The five constraints $f_i(1)=1$,
$f_\sigma^{\prime\prime}(1) = f_\omega^{\prime\prime}(1)$, and
$f_i^{\prime\prime}(0)=0$, reduce the number of independent parameters
to three. Two additional parameters in the isoscalar channel are
$g_\sigma(\rho_{\rm sat})$ and $g_\omega(\rho_{\rm sat})$. The functional
form of the density dependence of the $\rho$-meson coupling is
suggested by DB calculations of asymmetric nuclear matter~\cite{JL.98}
\begin{equation}
\label{drho}
g_{\rho}(\rho) = g_{\rho}(\rho_{\rm sat})~{\rm exp}
\left [-a_{\rho} (x-1)\right ]\;.
\end{equation}
The isovector channel is parameterized by $g_{\rho}(\rho_{\rm sat})$ and
$a_{\rho}$.

\begin{figure}
\includegraphics{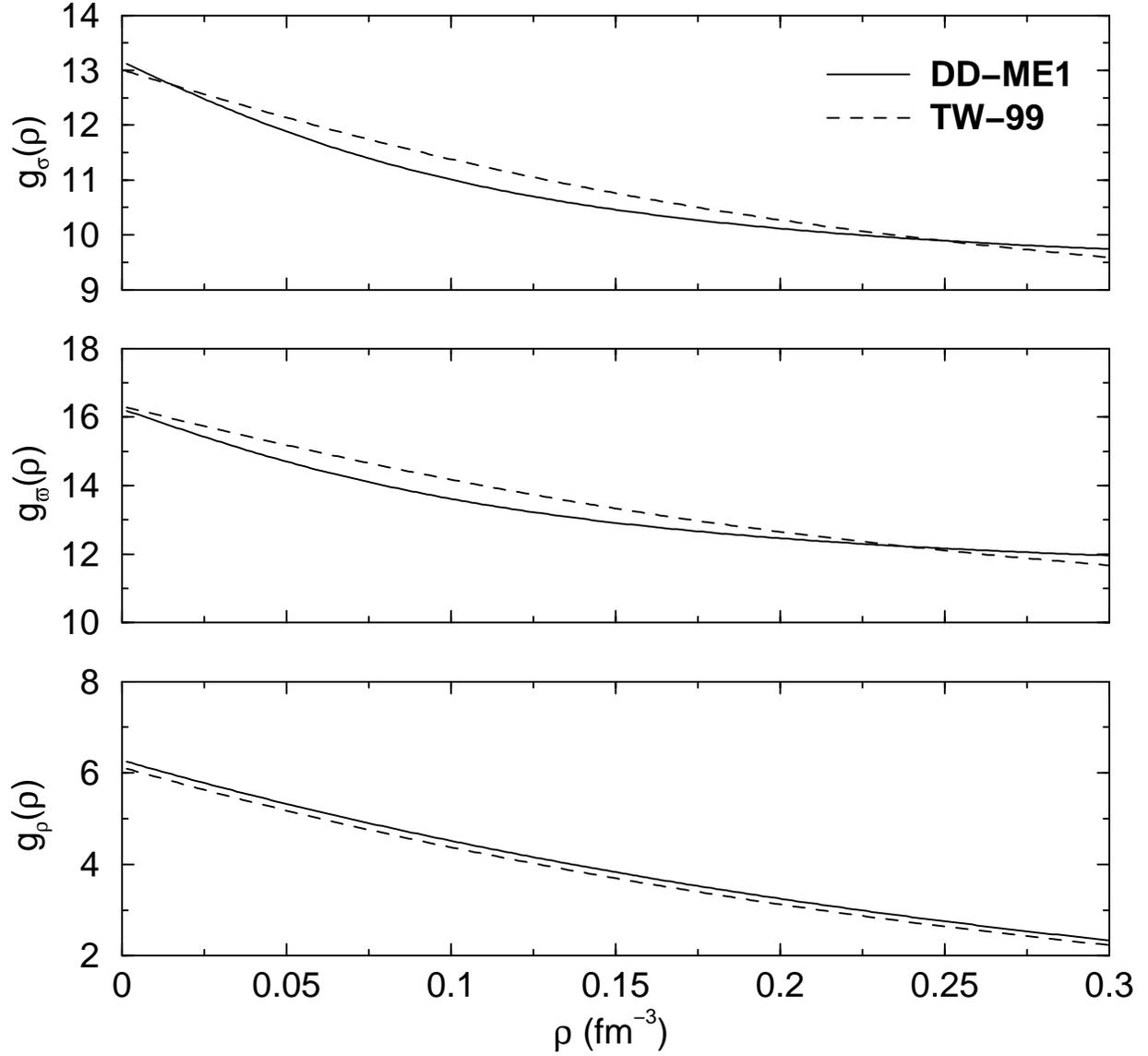}
\caption{\label{figA} Density dependence of the couplings of
the  $\sigma$-, $\omega$-, and $\rho$-meson. The result of the
present analysis (DD-ME1) is shown in comparison with the
parameters of the effective interaction TW-99~\protect\cite{TW.99}.}
\end{figure}

In Ref.~\cite{TW.99} the standard free values for the masses of
the $\omega$ and $\rho$ mesons were taken: $m_\omega = 783$ MeV
and $m_\rho = 763$ MeV. The mass of the $\sigma$ meson was fixed
to $m_\sigma = 550$ MeV, and the remaining seven independent
parameters were adjusted to nuclear matter properties (three
parameters) and to the binding energies of symmetric and
neutron-rich nuclei (four parameters). In this work the
density-dependent meson-nucleon couplings have been determined
using a somewhat different procedure. The parameters have been
adjusted simultaneously to properties of nuclear matter (see
Table~\ref{tab1}), and to binding energies, charge radii and
differences between neutron and proton radii of spherical nuclei
(see Table~\ref{tab3}). In addition to the seven coupling
parameters, the mass of the $\sigma$-meson has also been included
in the fitting procedure, i.e. we have used one more free
parameter with respect to the model of Ref.~\cite{TW.99}. For the
open shell nuclei in Table ~\ref{tab3}, pairing correlations have
been treated in the BCS approximation with empirical pairing gaps
(five-point formula). For nuclear matter the  "empirical" input
was: $E/A = -16$ MeV (5\%), $\rho_0 = 0.153$ fm$^{-3}$  (10\%), 
$K_0 = 250$ MeV (10\%), and $J = 33$ MeV (10\%). The values in  
parentheses correspond to the error bars used in the fitting procedure.
The binding energies of finite nuclei and the charge radii were
taken within an accuracy of 0.1\% and 0.2\%, respectively. 
Due to large experimental uncertainties, however, the error
bar used for the neutron skin was 5\%.  
After the solution of the self-consistent
equations, we subtract from the total binding energy the
microscopic estimate for the center-of-mass correction
\begin{equation}
\label{cms}
E_{cm} = - \frac{<P_{cm}^2>}{2Am} \;,
\end{equation}
where $P_{cm}$ is the total momentum of a nucleus with $A$ nucleons.
The resulting parameters of the density-dependent 
meson-exchange effective interaction
(DD-ME1) are displayed in Table~\ref{tab1}, in comparison to those of
Ref.~\cite{TW.99} (TW-99). We note two additional differences between
these models. In the relativistic mean-field model of
Ref.~\cite{TW.99} the center-of-mass correction (\ref{cms}) is
calculated by using the non-relativistic approximation for the
nucleon wave functions, and the Coulomb energy is corrected by
multiplying the vector self-energy of the proton with the factor
$(Z-1)/Z$. In Table~\ref{tab1} we first note that, even though
$m_\sigma$ is a free parameter in our model, the adjustment to nuclear
matter and to properties of finite spherical nuclei produces a value
that is very close to that of the TW-99 parameterization. The two
effective interactions display similar values for the meson-nucleon
coupling parameters at saturation density $g_i(\rho_{\rm sat})$
($i = \sigma,\omega,\rho$), as well as the value
of the parameter $a_{\rho}$, which determines the density-dependence
of the $\rho$-meson coupling (\ref{drho}).
The eight parameters, which characterize the density dependence
(\ref{func}) of the $\sigma$- and $\omega$-meson couplings
in DD-ME1, are very different from those of the TW-99 effective
interaction. The reason is, of course, that only three of these
parameters are independent. In our model $b_\sigma$ ,
$d_\sigma$ and $d_\omega$ are adjusted to properties of nuclear
matter and finite nuclei. Nevertheless, the overall density
dependence of the meson-nucleon vertex functions is very similar
in the two models, as shown in Fig.~\ref{figA} for the relevant
densities $\rho \leq 0.3$ fm$^{-3}$.

\begin{figure}
\includegraphics{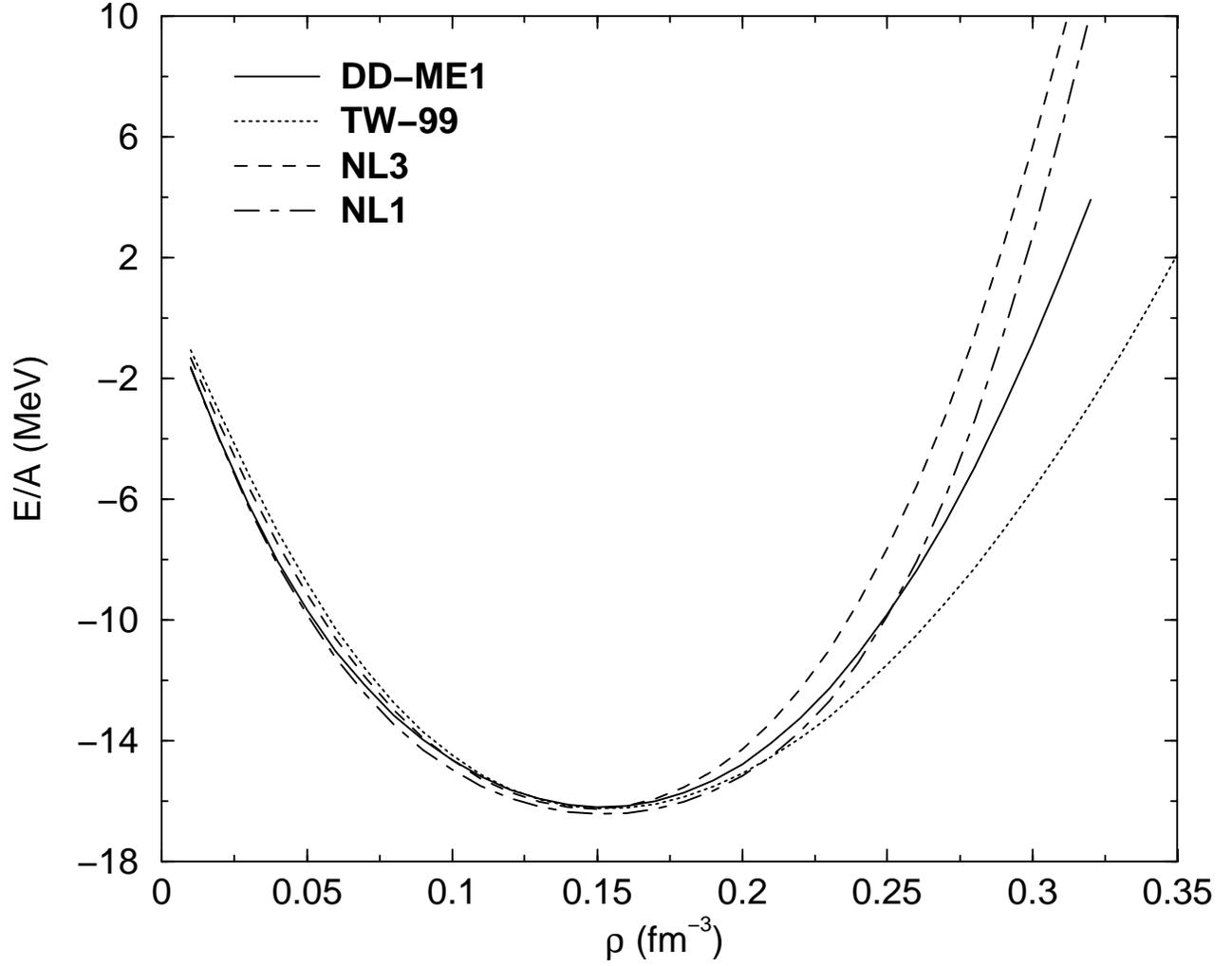}
\caption{\label{figB} Binding energy per nucleon for symmetric nuclear
matter as a function of the baryon density, calculated with the
density-dependent effective interactions DD-ME1 and TW-99~\protect\cite{TW.99},
and the non-linear parameter sets NL3~\protect\cite{LKR.97} and
NL1~\protect\cite{RRM.86}.}
\end{figure}

\begin{figure}
\includegraphics{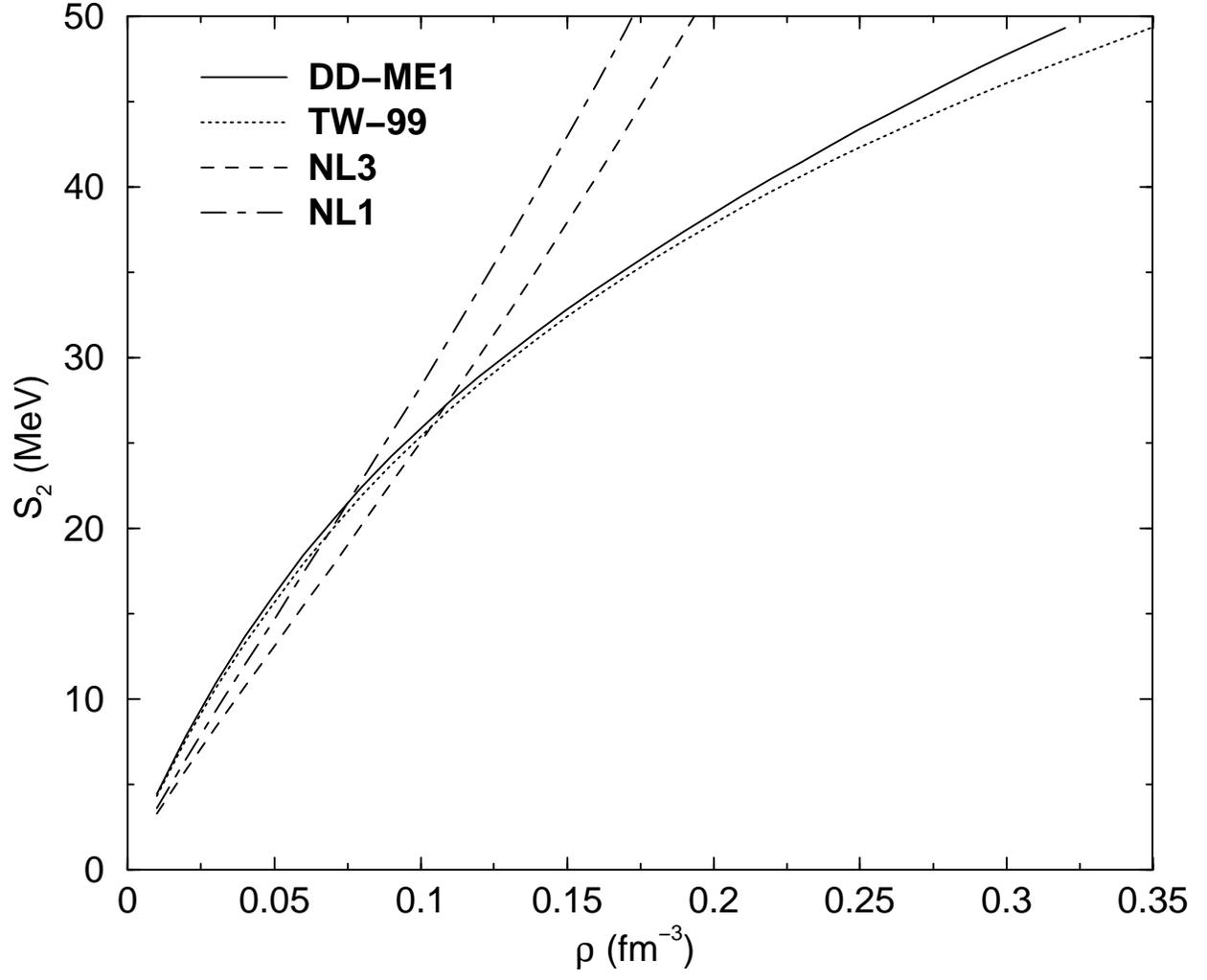}
\caption{\label{figC} $S_2(\rho)$ coefficient (\protect\ref{S2})
of the quadratic term of the energy per particle of asymmetric nuclear matter,
calculated with the effective interactions DD-ME1, TW-99~\protect\cite{TW.99},
NL3~\protect\cite{LKR.97}, and NL1~\protect\cite{RRM.86}.}
\end{figure}

\begin{figure}
\includegraphics{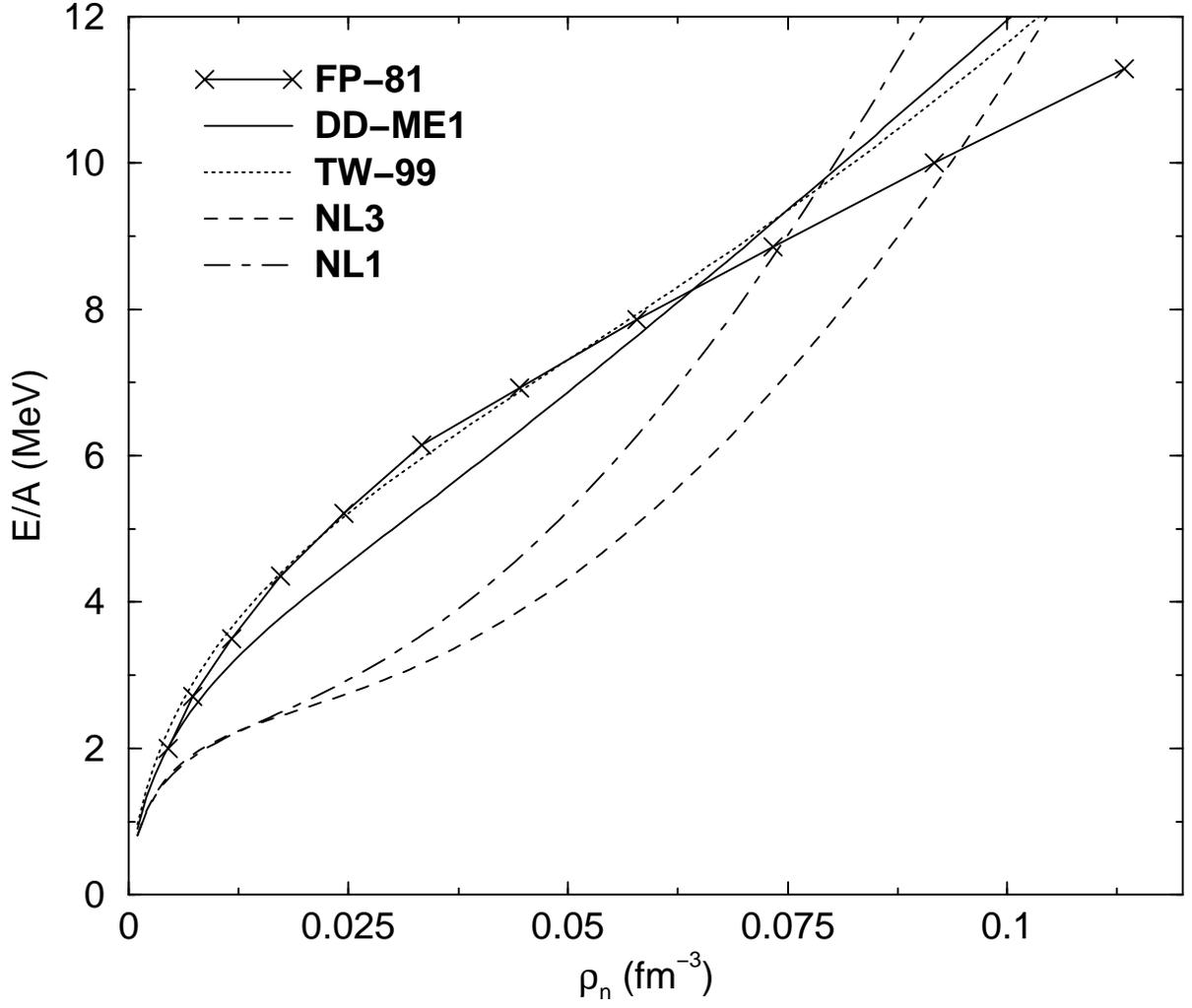}
\caption{\label{figD} Energy per particle of neutron matter as a
function of the neutron density. The results for the four
relativistic mean-field interactions DD-ME1, TW-99~\protect\cite{TW.99},
NL3~\protect\cite{LKR.97}, and NL1~\protect\cite{RRM.86} are
shown in comparison with the neutron matter equation of state of
Friedman and Pandharipande (FP-81)~\protect\cite{FP.81}.}
\end{figure}

Nuclear matter properties calculated with the DD-ME1 interaction
are illustrated in Table~\ref{tab2} and Figs.~\ref{figB}, \ref{figC},
and \ref{figD}. The results are shown in comparison with those obtained
with the density-dependent effective interaction TW-99~\cite{TW.99},
and with two standard non-linear parameter sets NL3~\cite{LKR.97} and
NL1~\cite{RRM.86}. The later non-linear effective interactions have been
used extensively in studies of nuclear structure phenomena over
the whole periodic table, from light nuclei to superheavy
elements. For symmetric nuclear matter all four interactions display
similar saturation densities (with NL3 at the low end), and binding
energies per nucleon (with NL1 at the high end). While three parameters
of TW-99 have been specifically adjusted to the values of
$\rho_{\rm sat}$, $E/A$ and the incompressibility $K_0$ shown in
Table~\ref{tab2}, the effective interactions DD-ME1, NL3 and NL1
have been simultaneously adjusted to properties of nuclear matter
and finite nuclei. The incompressibility modulus $K_0$ of the
two density-dependent interactions ($\approx 240-245$) MeV lies between
the values predicted by the non-linear interactions NL1 and NL3.
Calculations of the excitation energies of
isoscalar giant monopole resonances in
spherical nuclei in the time-dependent relativistic mean-field
framework~\cite{Vre.97}, and in the relativistic random-phase
approximation~\cite{Ma.01}, suggest that the nuclear matter
incompressibility modulus should be in the range $K_0 \approx 250 - 270$ MeV.
All four effective interactions display low values of the effective mass
$m^*$ that are, of course, necessary in order to reproduce the empirical
spin-orbit splittings in spherical nuclei. The equations of state
of symmetric nuclear matter are compared in Fig.~\ref{figB}. All four
binding energy per particle curves display a very similar dependence
on density below the saturation point $\rho_{\rm sat}$. Pronounced
differences show up at higher densities. In particular, the two
non-linear effective interactions NL1 and NL3 display a much steeper
increase of the binding energy, especially with respect to TW-99.
This is due to the fact that the $\omega$-meson, which dominates
at higher densities, enters linearly in NL1 and NL3, with constant
coupling. The equation of state calculated with the DD-ME1 interaction
shows an intermediate density dependence in this region, though
closer to NL1 and NL3.

The principal difference between the density-dependent effective
interactions DD-ME1 and TW-99 on one hand, and the non-linear
interactions NL3 and NL1 on the other, are the properties of
asymmetric matter. This is a very important point, because
different isovector properties in nuclear matter lead to
very different predictions for the properties of exotic nuclei with
extreme isospin values. The energy per particle of asymmetric nuclear
matter can be expanded about the equilibrium density $\rho_{\rm sat}$
in a Taylor series in $\rho$ and $\alpha$~\cite{Lee.98}
\begin{equation}
E(\rho,\alpha) = E(\rho,0) + S_2(\rho) \alpha^2 + S_4(\rho) \alpha^4 + \cdots
\label{taylor}
\end{equation}
where
\begin{equation}
\alpha \equiv \frac{N-Z}{N+Z} \;.
\end{equation}

\begin{equation}
E(\rho,0) = - a_v + \frac{K_0}{18 \rho_{\rm sat}^2}~
(\rho - \rho_{\rm sat})^2 + ...
\end{equation}
and
\begin{equation}
S_2(\rho) = a_4 + \frac{p_0}{\rho_{\rm sat}^2}~(\rho - \rho_{\rm sat}) +
\frac{\Delta K_0}{18 \rho_{\rm sat}^2}~ (\rho - \rho_{\rm sat})^2 + \cdots
\label{S2}
\end{equation}
\bigskip

The empirical value at saturation density $S_2(\rho_{\rm sat}) = a_4 =
30\pm 4$ MeV. The parameter $p_0$ defines the linear density dependence
of the symmetry energy, and $\Delta K_0$ is the correction to the
incompressibility. The contribution of the term $S_4(\rho) \alpha^4$
in (\ref{taylor}) is very small
in ordinary nuclei and the coefficient is not constrained in the mean-field
approximation. We first note that the non-linear effective interactions
NL1 and NL3 have a considerably larger value  $a_4$ of the symmetry energy
at saturation density. This is also true for other standard non-linear
parameter sets, and is due to the fact that the isovector channel of these
effective forces is parameterized by a single constant, the
density-independent $\rho$-meson
coupling $g_\rho$. With a single parameter in the isovector channel
it is not possible to reproduce simultaneously the empirical value
of $a_4$ and the masses of $N \neq Z$ nuclei. This only becomes possible
if a density dependence is included in the $\rho$-meson coupling,
as it is done in TW-99 and DD-ME1. In a recent analysis of neutron
radii in non-relativistic and covariant mean-field models~\cite{Fur.01},
Furnstahl has studied the linear correlation between the neutron skin
and the symmetry energy. In particular, he has shown that there is a
very strong linear correlation between the neutron skin thickness in
$^{208}$Pb and the individual parameters, which determine the
symmetry energy $S_2(\rho)$: $a_4$, $p_0$ and $\Delta K_0$. The empirical
value of $r_n - r_p$ in $^{208}$Pb ($0.20 \pm 0.04$ fm from proton
scattering data~\cite{SH.94}, and $0.19 \pm 0.09$ fm from the alpha scattering
excitation of the isovector giant dipole resonance~\cite{Kra.94}) places the
following constraints on the values of the parameters of the symmetry
energy: $a_4 \approx 30-34$ MeV, 2 Mev/fm$^3 \leq p_0 \leq$ 4 Mev/fm$^3$,
and $-200$ MeV $\leq \Delta K_0 \leq -50$ MeV. In Table~\ref{tab2} we
notice that, while these constraints are satisfied by the density-dependent
interactions DD-ME1 and TW-99, the parameters of the symmetry energy
of the non-linear interactions are systematically much larger. In particular,
$p_0$ is too large by a factor $\approx 2$, and the correction to the
incompressibility $\Delta K_0$ has even a wrong sign for the two
non-linear interactions. The qualitatively different density dependence
of the symmetry energy for the two classes of effective interactions is
also illustrated in Fig.~\ref{figC}, where we plot the coefficient
$S_2$ as a function of the baryon density. Due to the very
large value of $p_0$ and  the small absolute value of $\Delta K_0$,
for NL3 and NL1 $S_2$ displays an almost linear density dependence
of $\rho$. For the two density-dependent interactions, on the other
hand, the quadratic term of $S_2$ dominates, especially at densities
$\rho \geq 0.1$ fm$^{-3}$.

In Fig.~\ref{figD} we display the energy per particle of neutron matter as
a function of the neutron density. At low densities, which are relevant
for nuclear structure problems, the results for the four
relativistic mean-field interactions DD-ME1, TW-99, NL3 and NL1 are
shown in comparison with the microscopic many-body
neutron matter equation of state of
Friedman and Pandharipande~\cite{FP.81}. The later is well reproduced
by the density-dependent effective interactions, especially by
TW-99, while the two non-linear interactions NL3 and NL1 display
a qualitatively different neutron matter equation of state, even at
very low densities.

The density-dependent effective interaction DD-ME1 has been simultaneously
adjusted to nuclear matter properties and to ground-state properties
of the spherical nuclei shown in Table~\ref{tab3}. The calculated binding
energies, charge radii and differences between neutron and proton radii
are shown in comparison with available experimental data. The choice
of the isotopes of Sn and Pb for the fitting procedure was motivated by
the desire to construct an effective interaction that could be
applied in the description of long isotopic chains, including
exotic nuclei which lie very far from the valley of $\beta$-stability.
The overall agreement between the calculated quantities and experimental
data in Table~\ref{tab3} is very good.

One of the advantages of using the relativistic framework lies in the fact
that the effective single-nucleon spin-orbit potential arises naturally from
the Dirac-Lorentz structure of the effective Lagrangian. The
single-nucleon Hamiltonian
does not contain any adjustable parameter for the spin-orbit interaction.
In Table~\ref{tab4} we compare the energy spacings of spin-orbit partners
in the doubly closed-shell nuclei $^{16}$O, $^{40}$Ca, $^{48}$Ca,
$^{132}$Sn and $^{208}$Pb, with the values calculated with the
DD-ME1 interaction and with the prediction of the standard NL3
non-linear interaction. The experimental data are from Ref.~\cite{SO}.
We notice that, even though the values calculated with NL3 are already
in very good agreement with experimental data, a further improvement
is obtained with the DD-ME1 interaction, especially for the lighter
nuclei $^{16}$O, $^{40}$Ca and $^{48}$Ca.


\section{\label{secIV} Ground states of the ${\rm Sn}$ and
${\rm Pb}$ isotopes}
In Ref.~\cite{LVR.98} we have applied the
relativistic Hartree Bogoliubov (RHB) model in a detailed analysis
of ground-state properties of Ni and Sn isotopes.
The NL3 parameter set~\cite{LKR.97} was used for the effective
mean-field Lagrangian, and pairing correlations were described
by the pairing part of the finite range Gogny interaction
D1S~\cite{BGG.84}. Fully self-consistent RHB solutions were
calculated for the isotopic chains of Ni ($28\leq N\leq 50$)
and Sn ($50\leq N\leq 82$). Binding energies,
neutron separation energies, and proton and neutron $rms$ radii
were compared with experimental data. The reduction of the
spin-orbit potential with the increase of the number of neutrons
was studied, and the resulting
energy spacings between spin-orbit partners were
discussed, as well as pairing properties calculated
with a finite range effective interaction in the $pp$ channel.

In this section we test the new density-dependent meson-exchange
effective force DD-ME1 in comparison with the non-linear
interaction NL3. The RHB model is used to calculate ground-state
properties of Sn and Pb isotopes. Both NL3 and DD-ME1 mean-field
Lagrangians are employed for the $ph$ channel, and the pairing
part of the Gogny interaction D1S is used in the $pp$ channel.
This pairing interaction is a sum of two Gaussians
with finite range and properly chosen spin and isospin dependence.
The Gogny force has been very carefully adjusted to the pairing
properties of finite nuclei all over the periodic table.
Its basic advantage is the finite range, which automatically guarantees
a proper cut-off in momentum space. By comparing results of fully
self-consistent RHB calculations with experimental data, we
will show that the new effective interaction DD-ME1
provides an excellent description of ground-state properties
and, as compared with NL3, the isovector channel is considerably
improved.

\begin{figure}
\includegraphics{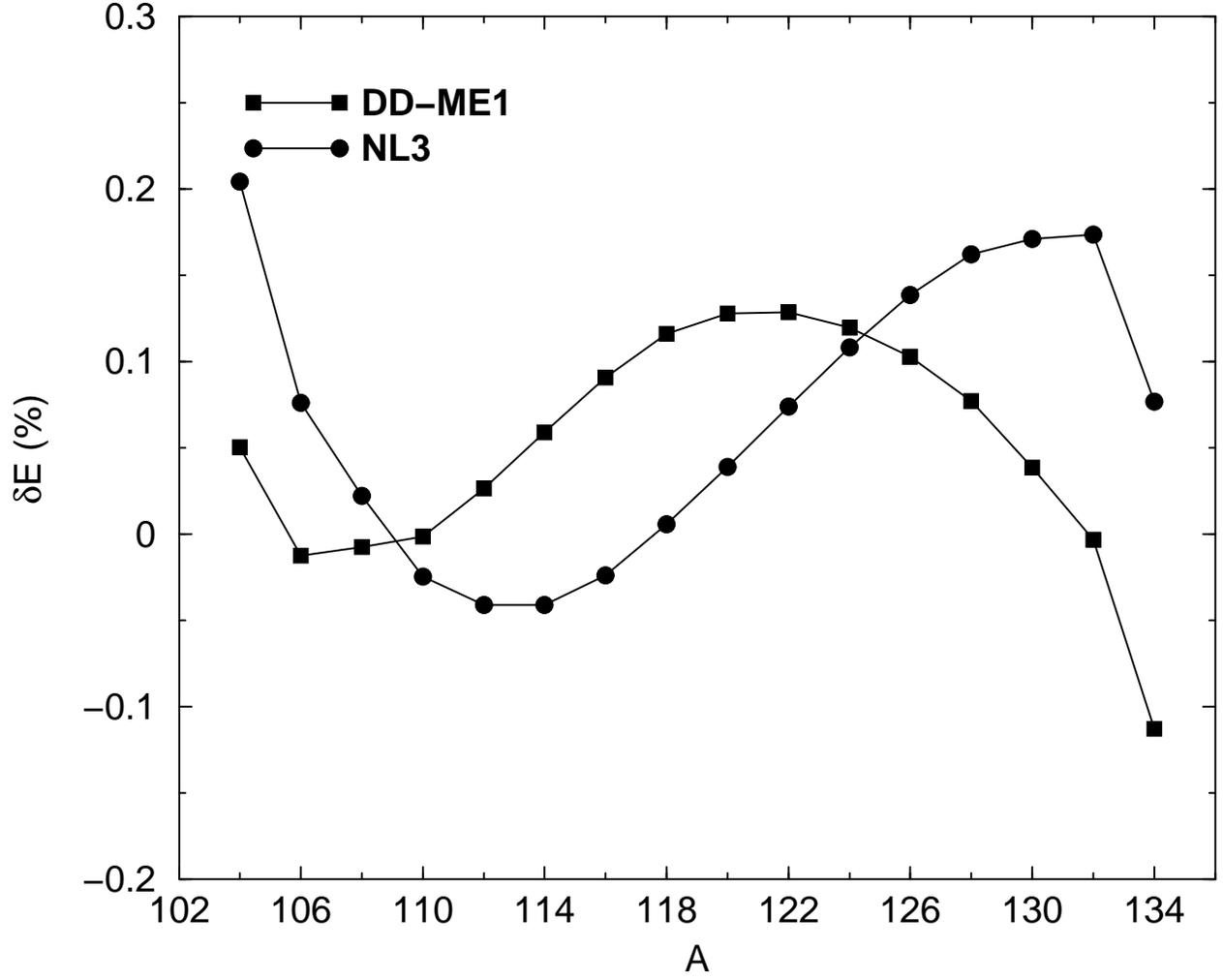}
\caption{\label{figE}The deviations (in percent) of the theoretical
masses of Sn isotopes, calculated in the RHB model with the
DD-ME1 and NL3 interactions,
from the empirical values~\protect\cite{AW.95}. The pairing part
of the Gogny interaction D1S has been used in the $pp$ channel of
the RHB model.}
\end{figure}

In Fig.~\ref{figE} we plot the deviations of the theoretical
masses of Sn isotopes, calculated in the RHB model with the
DD-ME1 and NL3 interactions, from the empirical values~\cite{AW.95}.
Both interactions display very good results over the entire
major shell $50 \leq N \leq 82$. For the new interaction DD-ME1,
in particular, only in few cases the absolute deviation
of the calculated mass exceeds 0.1\%.

\begin{figure}
\includegraphics{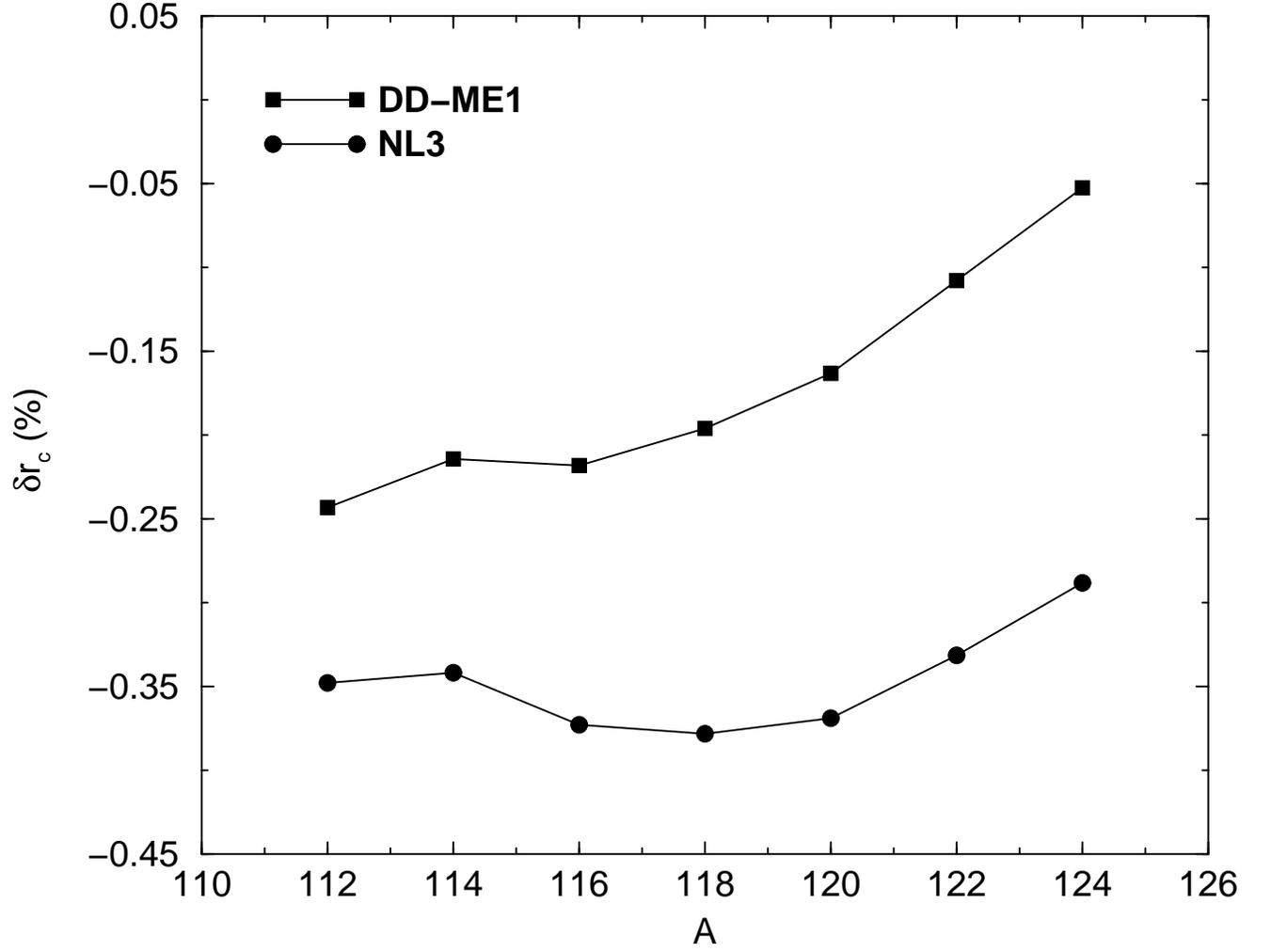}
\caption{\label{figF} The deviations (in percent) of the theoretical
charge radii of Sn isotopes, calculated in the RHB model with the
DD-ME1 and NL3 interactions,
from the experimental values~\protect\cite{CHARGE}.}
\end{figure}

The isotopic dependence of the difference between the theoretical
and experimental charge radii~\cite{CHARGE} of Sn nuclei is displayed
in Fig.~\ref{figF}. The charge radii calculated with both DD-ME1 and
NL3 interactions are systematically smaller than the experimental
values. The new density-dependent force, however, reduces the
deviations from the experimental radii by a factor $\approx 2$.
The parameters of DD-ME1 have been adjusted to the charge radii
of $^{112,116,124}$Sn, and the absolute deviations for these
nuclei can be compared in Table~\ref{tab3}.

\begin{figure}
\includegraphics{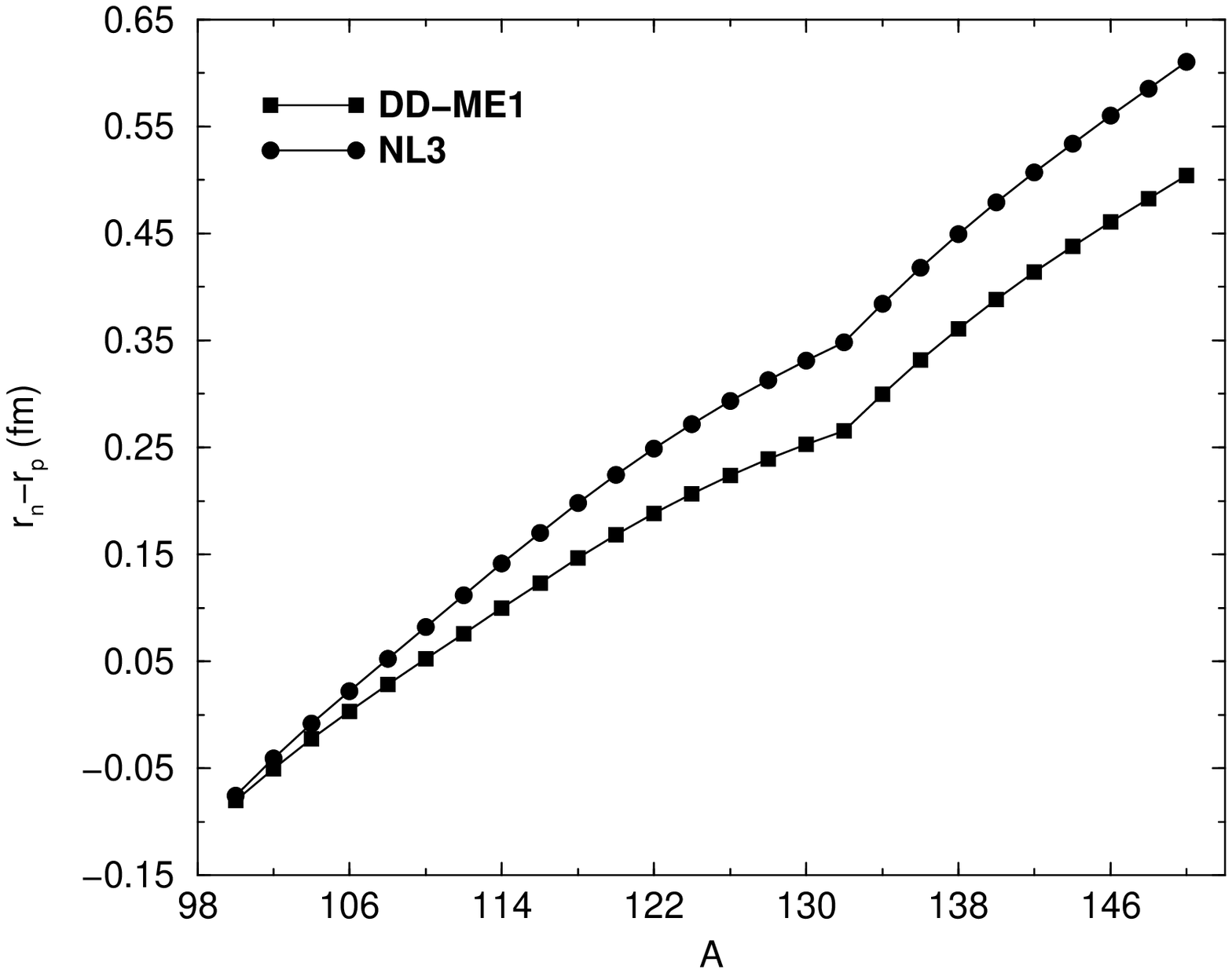}
\caption{\label{figG} Differences between neutron and proton radii
of ground-state distributions of Sn isotopes, calculated with the
DD-ME1 and NL3 effective interactions.}
\end{figure}

\begin{figure}
\includegraphics{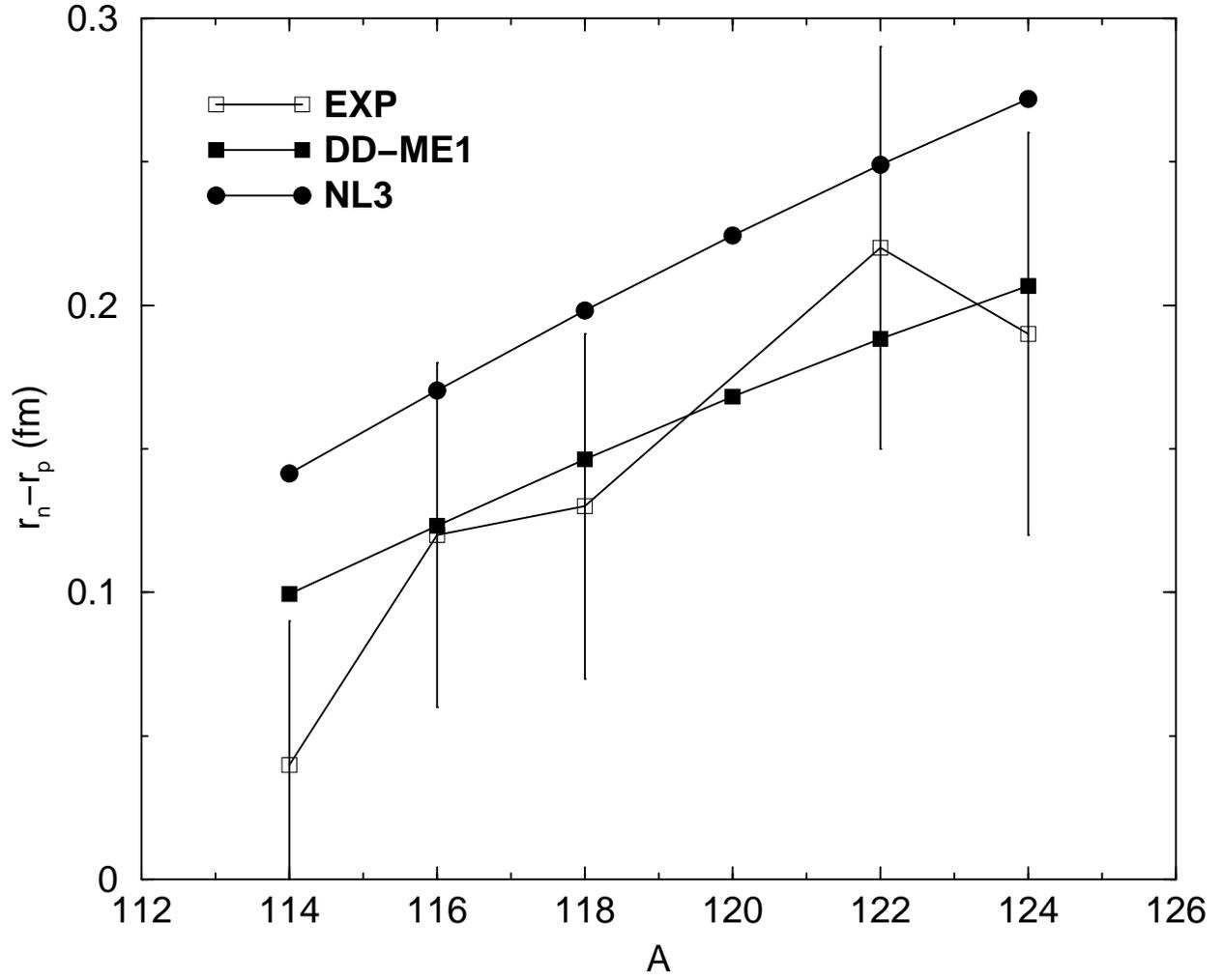}
\caption{\label{figH} DD-ME1 and NL3 predictions for the
differences between neutron and proton
$rms$ radii of Sn isotopes, compared with
experimental data from Ref.~\protect\cite{Kra.99}.}
\end{figure}

The calculated differences between radii of neutron and proton
ground-state distributions of Sn nuclei are shown in
Fig.~\ref{figG}. The non-linear interaction NL3 systematically
predicts larger values of $r_n - r_p$. This effect is even more
pronounced for the older parameter set NL1~\cite{SR.92}. The
difference between the values calculated with NL3 and DD-ME1
increases with the number of neutrons to about $0.1$ fm at $N=82$,
but then it remains practically constant for $N > 82$. The
calculated values of $r_n - r_p$ are compared with experimental
data~\cite{Kra.99} in Fig.~\ref{figH}. While both interactions
reproduce the isotopic trend of the experimental data, NL3
obviously overestimates the neutron skin.  The values calculated
with DD-ME1, on the other hand, are in excellent agreement with
the experimental data. This result presents a strong indication
that the isovector channel of the effective interaction DD-ME1 is
correctly parameterized.

\begin{figure}
\includegraphics{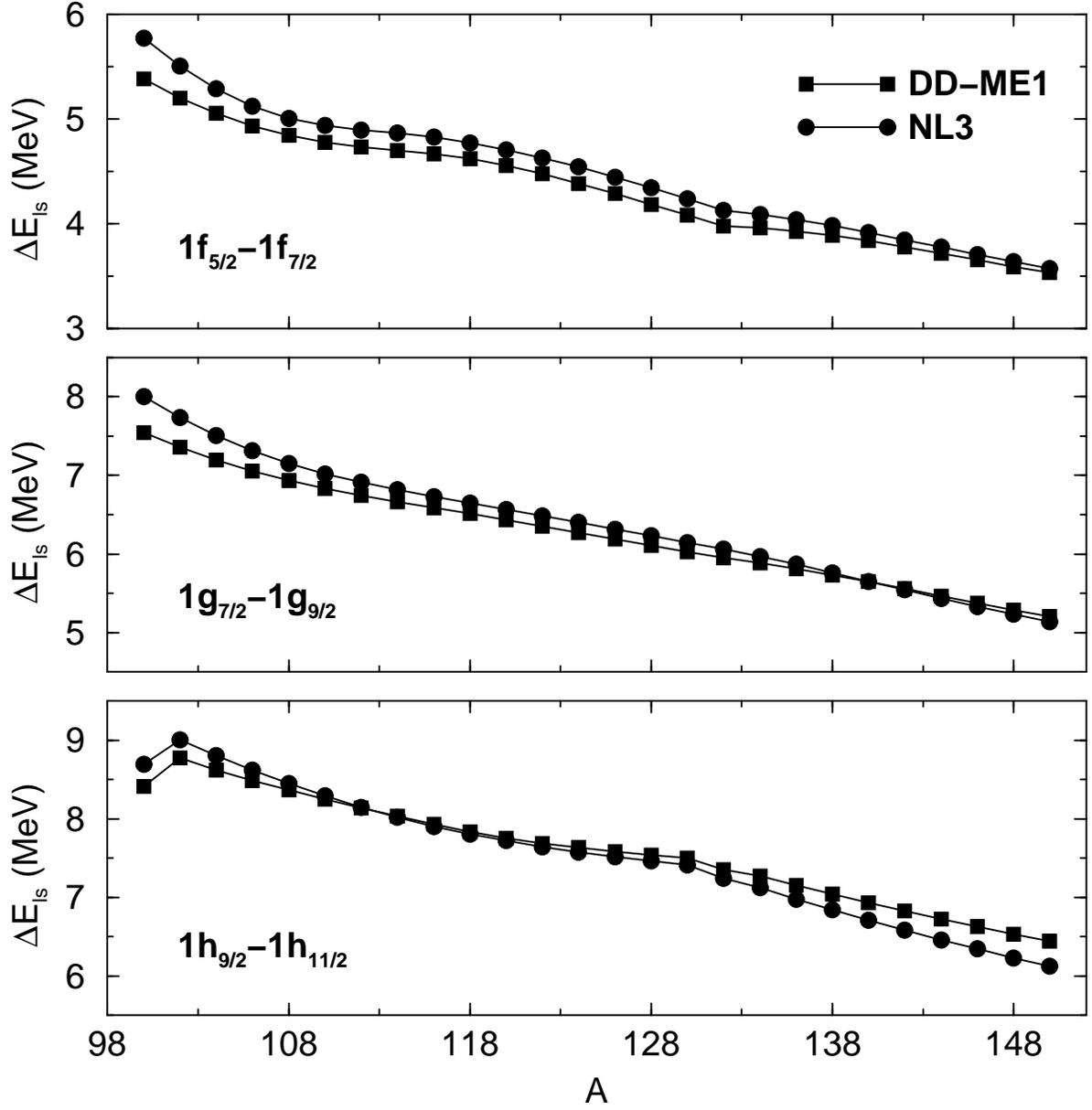}
\caption{\label{figI}Energy spacings between neutron
spin-orbit partner states in Sn isotopes, calculated in the RHB model
with the DD-ME1 and NL3 effective interactions.}
\end{figure}

In Refs.~\cite{LVR.97,LVR.98} it has been shown that the relativistic
mean-field framework predicts a strong reduction of the magnitude of
the spin-orbit term in the effective single nucleon potential of nuclei
with extreme isospin values. Starting from $T_z=0$ nuclei, and
increasing the number of neutrons or protons, the effective
spin-orbit interaction becomes weaker and this results in a reduction
of the energy spacings for spin-orbit partners. The spin-orbit
potential originates from the addition of two large fields:
the field of the vector mesons (short range repulsion), and
the scalar field of the sigma meson (intermediate
attraction). In the first order approximation, and
assuming spherical symmetry, the spin orbit term can be
written as
\begin{equation}
\label{so1}
V_{s.o.} = {1 \over r} {\partial \over \partial r} V_{ls}(r)\; ,
\end{equation}
where $V_{ls}$ is the spin-orbit potential
\begin{equation}
\label{so2}
V_{ls} = {m \over m_{eff}} (V-S)\; .
\end{equation}
V and S denote the repulsive vector and the attractive
scalar potentials, respectively.  $m_{eff}$ is the
effective mass
\begin{equation}
\label{so3}
m_{eff} = m - {1 \over 2} (V-S)\;.
\end{equation}
On the neutron-rich side the magnitude of the spin-orbit term $V_{s.o.}$
decreases as we add more neutrons, i.e. more units of isospin.
This is reflected in the energy spacings between the neutron
spin-orbit partner states
\begin{equation}
 \Delta E_{ls} = E_{n,l,j=l-1/2} - E_{n,l,j=l+1/2}\;.
\end{equation}
In Fig.~\ref{figI} we plot the energy spacings between neutron
spin-orbit partners in Sn isotopes, calculated in the RHB model
with the DD-ME1 and NL3 effective interactions. The calculated isotopic
dependence is almost identical. Both interactions predict a reduction
of the energy spacings between spin-orbit partners of $\approx 50\%$
in the interval $100 \leq A \leq 150$.

\begin{figure}
\includegraphics{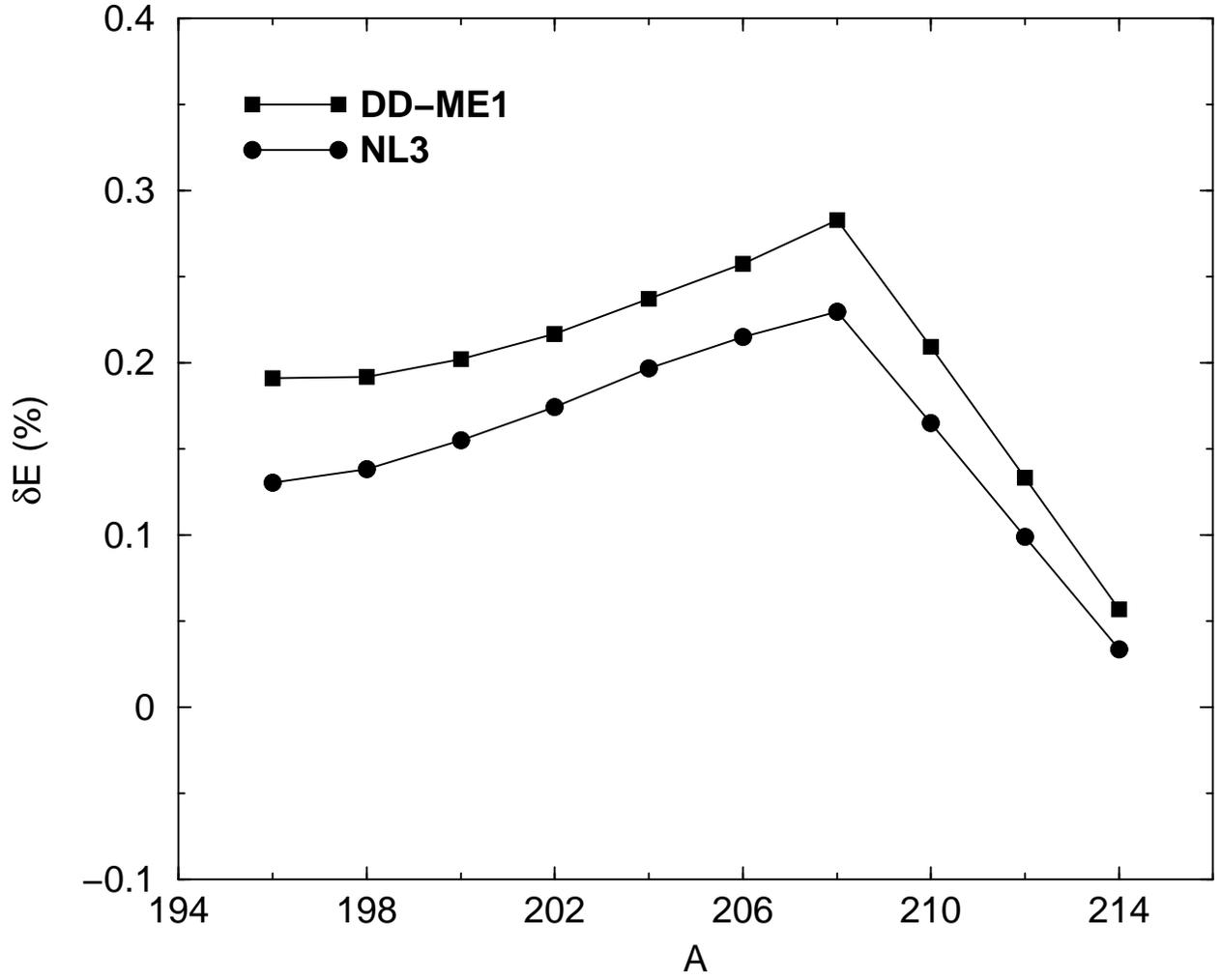}
\caption{\label{figJ} The deviations (in percent) of the theoretical
masses of Pb isotopes, calculated in the RHB model with the
DD-ME1 and NL3 interactions,
from the empirical values~\protect\cite{AW.95}.}
\end{figure}

In order to test the DD-ME1 effective interaction in the region of
heavy nuclei, we have calculated the Pb isotopes with $196 \leq A \leq 214$.
In Fig.~\ref{figJ} we display the deviations of the RHB theoretical
masses of Pb isotopes from the empirical values~\cite{AW.95}.
The accuracy of the binding energies calculated with DD-ME1 is
comparable to that obtained with the NL3 interaction. However,
in contrast to the case of Sn isotopes, the DD-ME1 interaction
systematically gives more binding as compared with NL3,
especially for $A < 208$.

\begin{figure}
\includegraphics{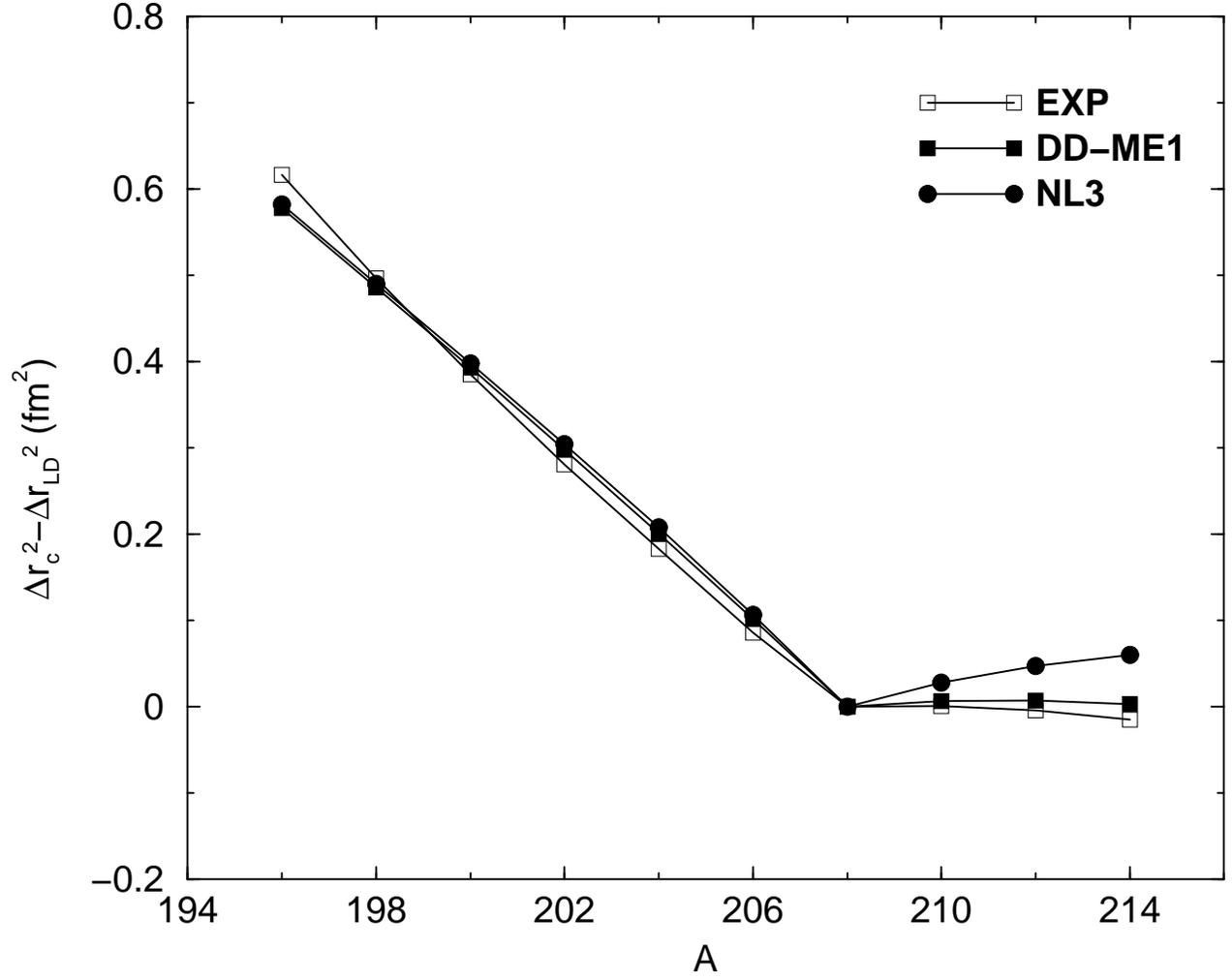}
\caption{\label{figK}Charge isotope shifts in even-A Pb isotopes.
The results of RHB
calculations with the DD-ME1 and NL3 effective interactions,
and with the Gogny D1S interaction in the pairing channel,
are compared with experimental data from Ref. ~\protect\cite{Rad}.}
\end{figure}

Due to the intrinsic isospin dependence of the effective
single-nucleon spin-orbit potential, the relativistic mean-field
models naturally reproduce the anomalous charge isotope
shifts~\cite{SLR.93}. The well known example
of the anomalous kink in the isotope shifts of
Pb isotopes is shown in Fig.~\ref{figK}. The results of RHB
calculations with the DD-ME1 and NL3 effective interactions,
and with the Gogny D1S interaction in the pairing channel,
are compared with experimental data from Ref.~\cite{Rad}.
Both interactions reproduce the general trend
of isotope shifts and the kink at $^{208}$Pb. The effect is
however, too strong with NL3. The experimental data
are better described by the DD-ME1 interaction.

\begin{figure}
\includegraphics{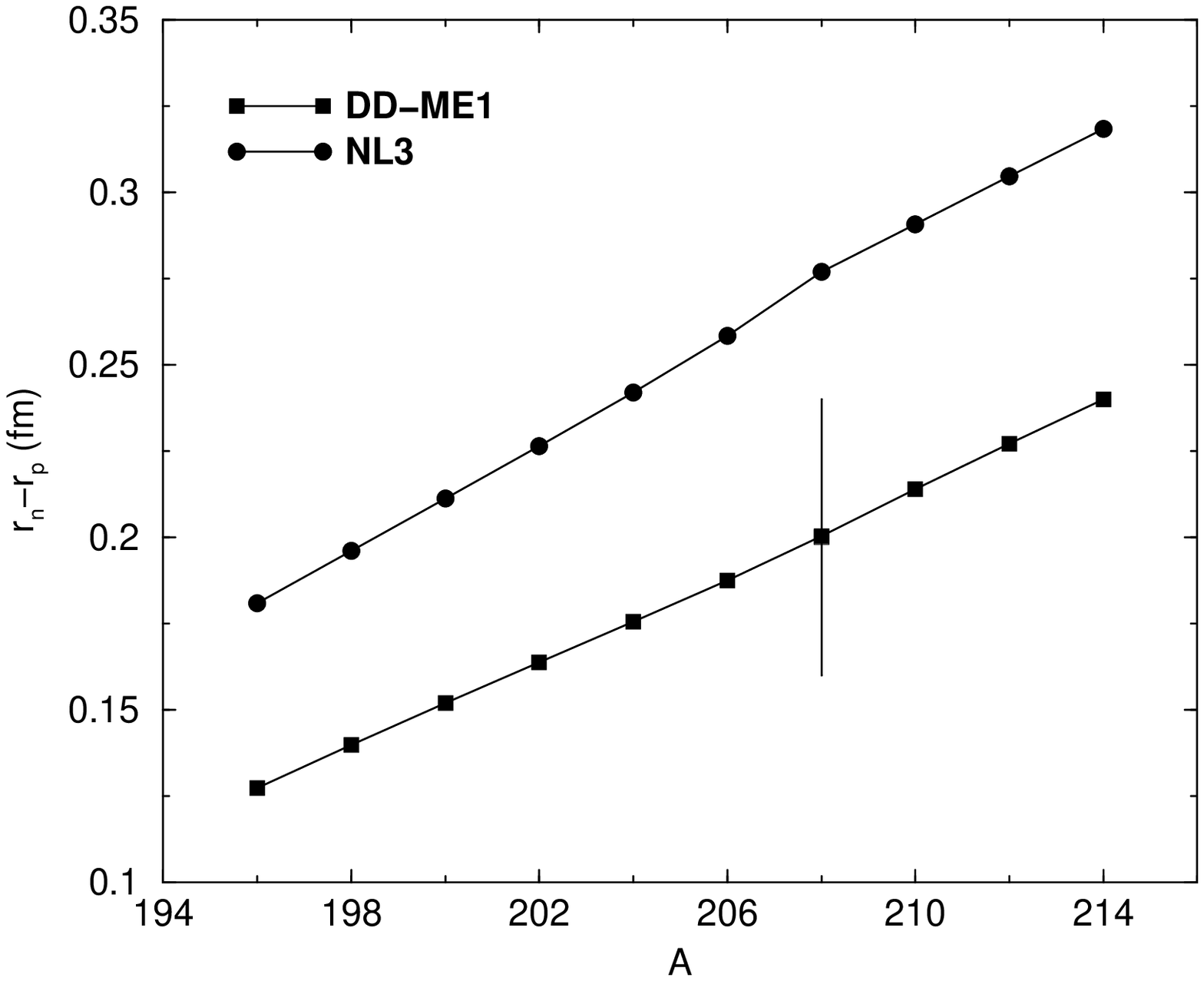}
\caption{\label{figL} Differences between neutron and proton radii
of ground-state distributions of Pb isotopes, calculated with the
DD-ME1 and NL3 effective interactions.}
\end{figure}

Finally, in Fig.~\ref{figL} we display the differences between
radii of neutron and proton ground-state distributions of Pb
isotopes, calculated with the DD-ME1 and NL3 effective
interactions. Similar to the case of Sn isotopes, DD-ME1
systematically predicts much smaller values for $r_n-r_p$, in
better agreement with available experimental data. The
experimental values of $r_n - r_p$ in $^{208}$Pb are: $0.20 \pm
0.04$ fm deduced from proton scattering data~\cite{SH.94}, and
$0.19 \pm 0.09$ fm deduced from the alpha scattering excitation of
the isovector giant dipole resonance~\cite{Kra.94}. In a recent
analysis of intermediate energy nucleon elastic scattering data,
and correlated with analyses of electron scattering data, a value
$\approx 0.17$ fm was suggested for the neutron skin thickness in
$^{208}$Pb~\cite{Kar.02}. As it has been emphasized in a recent
analysis of neutron radii in mean-field models~\cite{Fur.01}, the
value of $r_n - r_p$ in $^{208}$Pb is crucial for constraining the
isovector channel of effective interactions in the mean-field
approach, both in non-relativistic and covariant models.

\section{\label{secV}Parity-violating elastic electron scattering
and neutron density distributions}
Data on neutron radii and neutron density distributions provide not only basic
nuclear structure information, but they also place
additional constraints on effective interactions used in nuclear
models. Potentially, a very accurate experimental method for the
determination of neutron densities is the elastic scattering of longitudinally
polarized electrons on nuclei. The parity-violating
asymmetry parameter, defined as the difference between cross sections
for the scattering of right- and left-handed longitudinally
polarized electrons, produces direct information on the
Fourier transform of the neutron density~\cite{DDS.89}.
A recent extensive analysis of possible parity-violating
measurements of neutron densities, their theoretical interpretation, and
applications can be found in Refs.~\cite{Hor.98,Hor.99}

In Ref.~\cite{Vre.00} we have studied parity-violating elastic electron
scattering on ground-state densities of neutron-rich nuclei that
were calculated in the RHB model with the NL3 + Gogny D1S interaction.
For the elastic scattering of 850 MeV electrons on these nuclei, the
parity-violating asymmetry parameters were calculated
using a relativistic optical model with inclusion of
Coulomb distortion effects. The asymmetry parameters
for chains of isotopes were compared, and their relation
to the Fourier transforms of neutron densities was studied.
In this work we have shown that the new density-dependent effective
interaction DD-ME1 predicts ground-state neutron density distributions
that are in much better agreement with experimental data. Thus, in this
section we include an analysis of parity-violating elastic electron
scattering on $^{208}$Pb and on those Sn isotopes for which there
are data on $r_n - r_p$  values.

We consider elastic
electron scattering on a spin-zero nucleus, i.e. on the potential
\begin{equation}
\hat V(r) = V(r) + \gamma_5 A(r)\;,
\label{pot1}
\end{equation}
where V(r) is the Coulomb potential, and A(r) results from the
weak neutral current amplitude
\begin{equation}
A(r)= {G_F\over 2^{3/2}} \rho_W(r)\;.
\label{pot2}
\end{equation}
The weak charge density is defined
\begin{equation}
\rho_W(r)=\int d^3r^\prime G_E(|{\bf r}-{\bf r}^\prime|)
[-\rho_n(r^\prime) + (1-4{\rm sin}^2\Theta_W)\rho_p(r^\prime)]\;,
\label{rhoW}
\end{equation}
where $\rho_n$ and $\rho_p$ are point neutron and proton densities and the
electric form factor of the proton is $G_E(r)\approx {\Lambda^3
\over 8\pi}e^{-\Lambda r}$ with $\Lambda=4.27$ fm$^{-1}$,
sin$^2\Theta_W=0.23$ for the Weinberg angle.

In the limit of vanishing electron mass, the electron spinor $\Psi$
defines the helicity states
\begin{equation}
\Psi_\pm={1\over 2}(1\pm\gamma_5)\Psi\;,
\end{equation}
which satisfy the Dirac equation
\begin{equation}
[{\bm \alpha}\cdot {\bm p} + V_\pm(r)]\Psi_\pm = E \Psi_\pm\;,
\end{equation}
with
\begin{equation}
V_\pm(r) = V(r) \pm A(r).
\end{equation}
The parity-violating asymmetry $A_l$, or helicity asymmetry, is
defined
\begin{equation}
A_l={d\sigma_+/d\Omega - d\sigma_-/d\Omega \over
       d\sigma_+/d\Omega + d\sigma_-/d\Omega}\;,
\label{AP}
\end{equation}
where $+(-)$ refers to the elastic scattering on the potential $V_\pm(r)$.
This difference arises from the interference of one-photon and
$Z^0$ exchange.

Starting from the relativistic Hartree-Bogoliubov solutions for
the self-consistent ground states, the charge and weak
densities are calculated by folding the point proton and
neutron densities. These densities define the Coulomb and
weak potentials in the Dirac equation for the massless electron.
The partial wave Dirac equation is solved with the inclusion
of Coulomb distortion effects, and the cross sections for
positive and negative helicity electron states are calculated.
The parity-violating asymmetry parameters are plotted as
functions of the scattering angle $\theta$, or the
momentum transfer $q$, and they are compared with the
Fourier transforms of the neutron density distributions.

\begin{figure}
\includegraphics{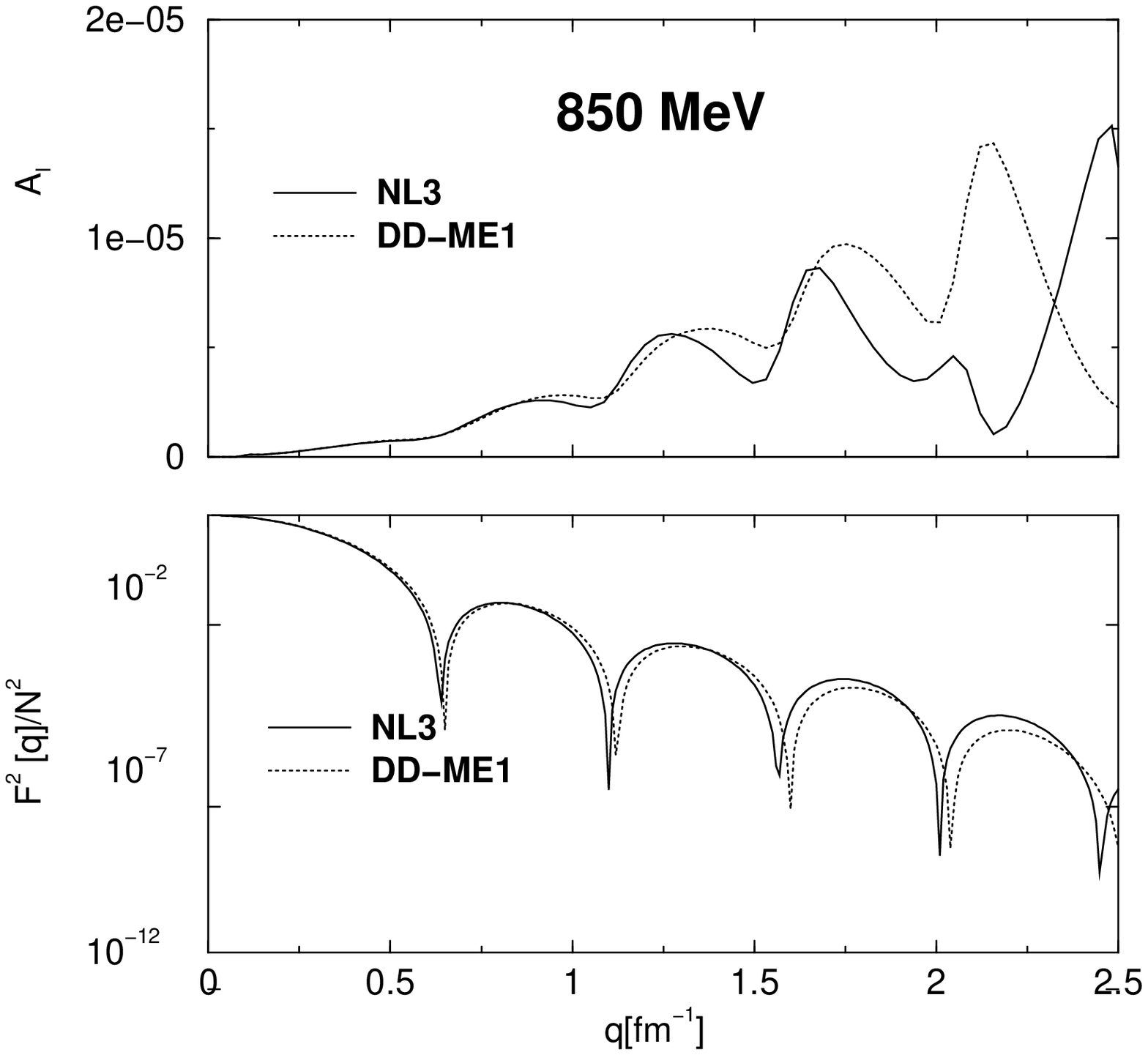}
\caption{\label{figM} Parity-violating asymmetry parameters $A_l$
(upper panel) and squares of normalized Fourier transforms
of neutron densities (lower panel), as functions of the
momentum transfer $q$, for elastic scattering from
$^{208}$Pb at 850 MeV.}
\end{figure}

In Fig.~\ref{figM} we plot the parity-violating
asymmetry parameters $A_l$ for elastic electron scattering from
$^{208}$Pb at 850 MeV, as functions of the
momentum transfer $q = 2 E sin{\theta /2}$, and compare
them with the squares of the Fourier transforms of the neutron
densities
\begin{equation}
F(q) = {4\pi \over q} \int dr~r^2 j_0(qr) \rho_n(r)\;.
\end{equation}
The solid and dotted curves correspond to RHB neutron ground state
densities calculated with the NL3 and DD-ME1 effective Lagrangians,
respectively. The asymmetries $A_l$ are of order of $\leq 10^{-5}$ and
increase with the momentum transfer $q$.
We notice that, even though the values
of $r_n - r_p$ calculated with the two interactions differ by
$\approx 0.07$ fm, the differences between the calculated
asymmetry parameters $A_l$ become more pronounced only for
$q > 1$ fm$^{-1}$.

\begin{figure}
\includegraphics{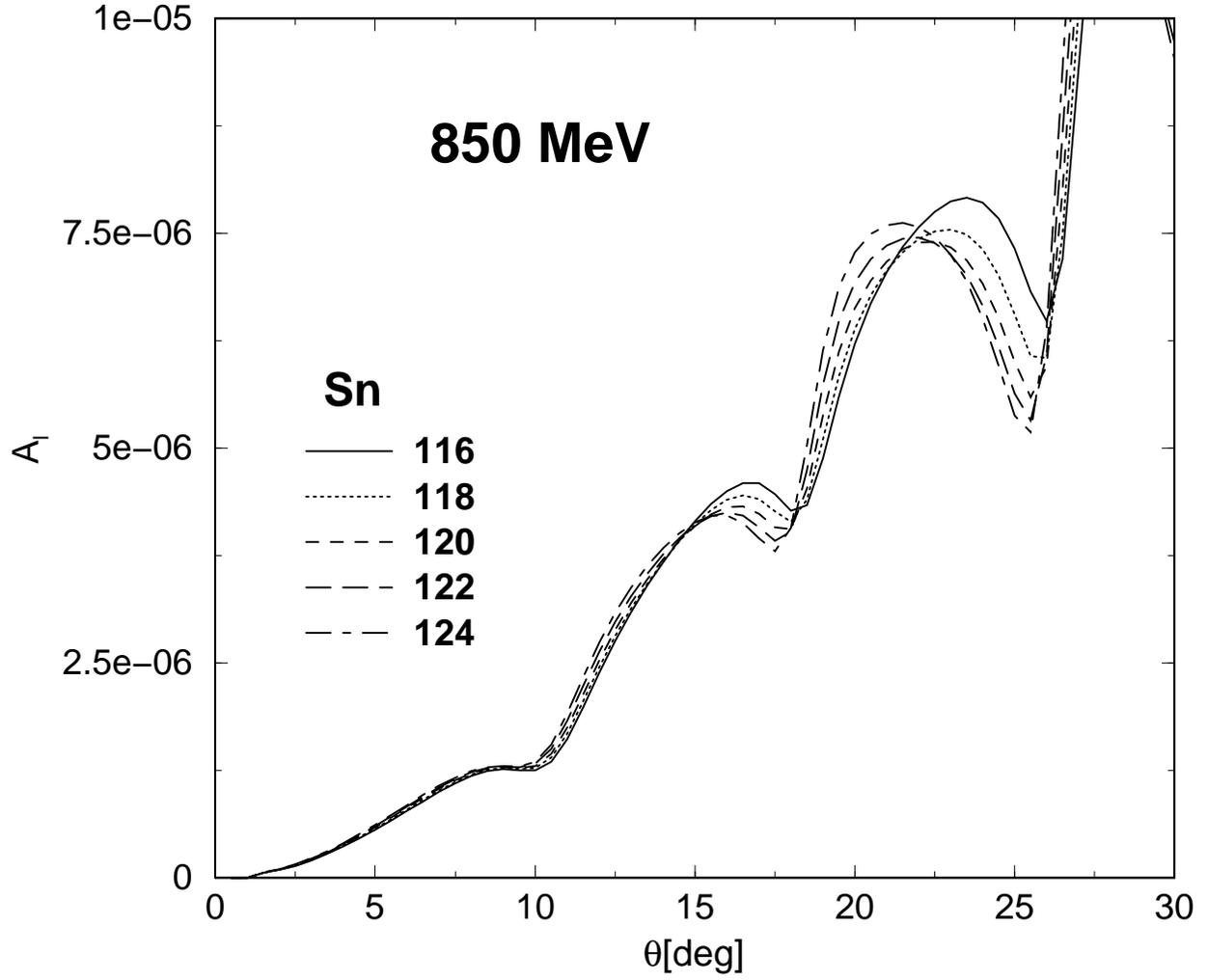}
\caption{\label{figN} Parity-violating asymmetry parameters $A_l$ for
elastic scattering from $^{116-124}$Sn at 850 MeV,
as functions of the scattering angle $\theta$.}
\end{figure}

\begin{figure}
\includegraphics{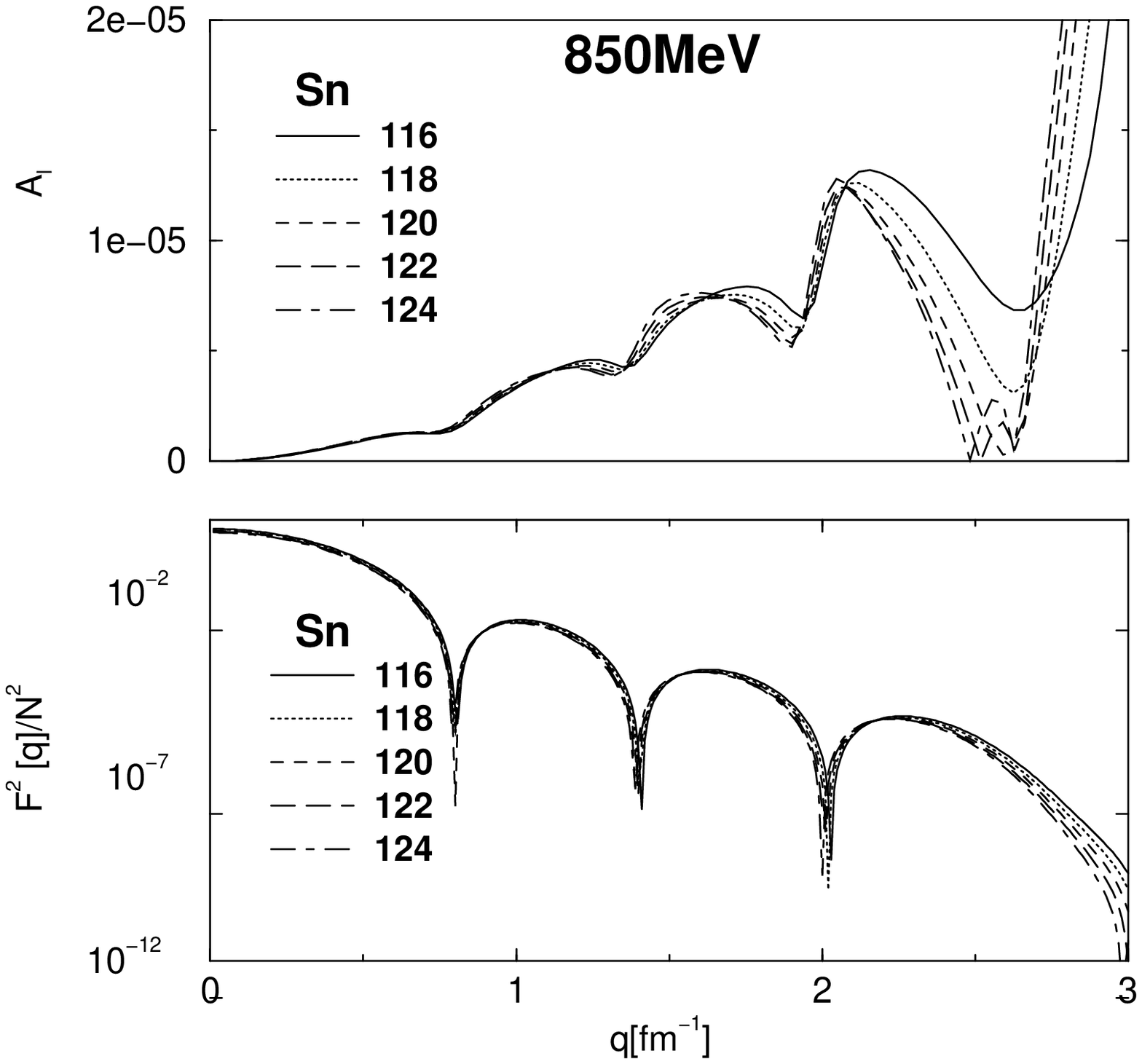}
\caption{\label{figO} Parity-violating asymmetry parameters $A_l$
(upper panel) and squares of normalized Fourier transforms
of neutron densities (lower panel), as functions of the
momentum transfer $q$, for elastic scattering from
$^{116-124}$Sn at 850 MeV.}
\end{figure}

The parity-violating asymmetry parameters $A_l$ for
elastic scattering from even-A isotopes $^{116-124}$Sn at 850 MeV,
as functions of the scattering angle $\theta$, are shown in
Fig.~\ref{figN}. They correspond to the ground-state densities
calculated with the DD-ME1 interaction. In Ref.~\cite{Vre.00} we have
discussed the sensitivity of the asymmetry parameters to the
formation of the neutron skin. For the Sn isotopes this effect is
illustrated in Fig.~\ref{figG}, where the
calculated differences between neutron and proton radii of
ground-state distributions are shown in comparison with experimental data.
The corresponding asymmetry parameters $\leq 10^{-5}$ in
Fig.~\ref{figN} display somewhat more pronounced differences between
neighboring isotopes only for $\theta > 15^o$. Finally, in Fig.~\ref{figO}
the calculated asymmetry parameters, as functions of the momentum transfer,
are compared with the squares of the Fourier transforms of the neutron
densities for $^{116-124}$Sn. Obviously a resolution better than
$\leq 10^{-6}$ for the asymmetry parameters is necessary in order
to obtain useful informations on neutron density distributions from
parity-violating elastic electron scattering. This resolution might be
already available at existing experimental facilities~\cite{Hor.99}.
For the Sn isotopes, in particular,
the differences between neighboring isotopes are only seen for
$q \geq 1.5 $ fm$^{-1}$. We should also mention that the magnitude of the
calculated asymmetry parameters depends, of course, on the electron
energy. For electron energies below 500 MeV the asymmetry
parameters are small~\cite{Vre.00}, while above 1 GeV the
approximation of elastic scattering on continuous charge and
weak densities is not valid any more, and the structure of
individual nucleons becomes important.

\section{\label{secVI}Summary and conclusions}

In the last couple of years
the relativistic Hartree-Bogoliubov (RHB) model has been very
successfully applied in the description of a variety of nuclear
structure phenomena. With the standard non-linear meson-exchange
relativistic mean-field effective interactions in the $ph$-channel,
however, the predictive power of the RHB model is somewhat limited,
especially for isovector properties of exotic nuclei far from
$\beta$-stability. We have tried to overcome these limitations
by extending the RHB model to include density-dependent meson-nucleon
couplings. The particular implementation of the model presented
in this work is based on the density dependent relativistic
hadron field (DDRH) theory~\cite{FL.95,FLW.95}.
The effective Lagrangian in the $ph$-channel is
characterized by a density dependence of the $\sigma$,
$\omega$ and $\rho$ meson-nucleon vertex functions.
The single-nucleon Dirac equation includes the
additional rearrangement self-energies that result from the
variation of the vertex functionals with respect to the baryon
field operators, and which are essential for the energy-momentum
conservation and the thermodynamical consistency of the model.
In this work we have used the phenomenological
density functional forms of the meson-nucleon coupling vertices
introduced in the DDRH framework by Typel and Wolter ~\cite{TW.99}.
The parameters of the new effective interaction DD-ME1 have been
determined by a multiparameter fit constrained by properties of
nuclear matter and by a set of experimental data on ground-state
properties of spherical nuclei. Pairing correlations in the
$pp$-channel of the RHB model are described by
the pairing part of the finite range Gogny interaction.

Properties of symmetric and asymmetric nuclear matter calculated with
the new density-dependent effective interaction DD-ME1 have been
compared to those obtained with the effective interaction of
Typel and Wolter~\cite{TW.99}, and with two standard non-linear
parameter sets sets NL3~\cite{LKR.97} and NL1~\cite{RRM.86}.
It has been shown that the density dependent meson-nucleon couplings
improve the behavior of the nuclear matter equation of state at
higher densities and reproduce the empirical value of the asymmetry
energy at saturation density. The properties of asymmetric nuclear
matter are much better described by the two density dependent interactions
and, in contrast to the non-linear NL3 and NL1 forces, these interactions
reproduce the microscopic many-body neutron matter equation of state
of Friedman and Pandharipande.

The RHB model with the density-dependent interaction DD-ME1 in the
$ph$-channel, and with the finite range Gogny interaction D1S in
the $pp$-channel, has been tested in the calculation of
ground-state properties of Sn and Pb isotopes. Results of fully
self-consistent RHB calculations of binding energies, charge
radii, differences between neutron and proton radii, spin-orbit
splittings, performed with the interaction DD-ME1 and with the
non-linear interaction NL3, have been compared with available
experimental data. While both interactions predict nuclear masses
with the same level of accuracy (absolute deviations $\approx$
$0.1-0.2$\%), the improved isovector properties of DD-ME1 result
in a better description of charge radii, and especially the
calculated values of $r_n - r_p$ are in much better agreement with
experimental data. The correct description of the data on
differences of the radii of neutron and proton ground-state
distributions, on neutron radii and neutron density distributions
is very important for studies of new phenomena in exotic nuclei
far from $\beta$-stability (neutron skin, neutron halo, pygmy
isovector dipole resonances), for astrophysical applications
(properties of neutron stars, neutron capture rates), and for a
theoretical interpretation of measurements of parity
nonconservation effects in atomic systems (tests of the Standard
model of electroweak interactions)~\cite{VLR.00}. In principle,
very accurate data on neutron density distributions could be
obtained from the elastic scattering of longitudinally polarized
electrons on nuclei. Using the ground-state densities calculated
with the DD-ME1 interaction in the RHB model, we have performed an
analysis of parity-violating elastic electron scattering on
$^{208}$Pb and on those Sn isotopes for which there are
experimental data on $r_n - r_p$. For the elastic scattering of
850 MeV electrons on these nuclei, the parity-violating asymmetry
parameters have been calculated using a relativistic optical model
with inclusion of Coulomb distortion effects, and related to the
Fourier transforms of the neutron density distributions.

The RHB model with density-dependent meson-nucleon couplings represents
a significant improvement in the relativistic mean-field description
of the nuclear many-body problem and, in particular, of exotic nuclei
far from $\beta$-stability. The improved isovector properties of the
effective interaction in the $ph$-channel on one hand, and the
unified description of mean-field and pairing correlations in the
Hartree-Bogoliubov framework on the other, offer a unique
possibility for accurate studies of nuclei with extreme ground-state
isospin values and with Fermi levels close to the particle continuum.
Particularly interesting will be studies of deformed nuclei for
which unusual shape coexistence phenomena, and even very different
proton and neutron ground-state deformations, are expected far
from stability and close to the drip-lines. Isovector ground-state
deformations could also give rise to exotic modes of low-energy isovector
collective excitations. We have already started with density-dependent
RHB calculations of deformed nuclei, and work is also in progress
on the description of collective excitations in the framework of
relativistic RPA/QRPA with density-dependent interactions.

\bigskip
\bigskip
\leftline{\bf ACKNOWLEDGMENTS}

This work has been supported in part by the
Bundesministerium f\"ur Bildung und Forschung under
project 06 TM 979, and by the Gesellschaft f\" ur
Schwerionenforschung (GSI) Darmstadt.

=========================================================================

\newpage
\begin {table}[]
\begin {center}
\caption {The effective interaction DD-ME1. The masses and meson-nucleon
couplings are shown in comparison with the parameters of the
density-dependent mean-field model of Ref.~\protect\cite{TW.99} (TW-99).}
\begin {tabular}{ccc}
          &          DD-ME1     &   TW-99 \\ \hline
\hline
$m_{\sigma}$             & {549.5255 } & {550.0000 }  \\
$m_{\omega}$             & {783.0000 } & {783.0000 }  \\
$m_{\rho}  $             & {763.0000 } & {763.0000 }  \\
$g_{\sigma}(\rho_{sat})$ & { 10.4434 } & { 10.7285 }  \\
$g_{\omega}(\rho_{sat})$ & { 12.8939 } & { 13.2902 }   \\
$g_{\rho}(\rho_{sat})  $ & {  3.8053 } & {  3.6610 }  \\
$a_{\sigma}$             & {  1.3854 } & {  1.3655 }\\
$b_{\sigma}$             & {  0.9781 } & {  0.2261 }  \\
$c_{\sigma}$             & {  1.5342 } & {  0.4097 }\\
$d_{\sigma}$             & {  0.4661 } & {  0.9020 }\\
$a_{\omega}$             & {  1.3879 } & {  1.4025 }\\
$b_{\omega}$             & {  0.8525 } & {  0.1726 }\\
$c_{\omega}$             & {  1.3566 } & {  0.3443 }\\
$d_{\omega}$             & {  0.4957 } & {  0.9840 }\\
$a_{\rho}$               & {  0.5008 } & {  0.515  }

\end {tabular}
\label{tab1}
\end{center}
\end{table}
\begin {table}[]
\caption{Nuclear matter properties calculated with the
density-dependent effective interactions DD-ME1 and TW-99~\protect\cite{TW.99},
and the non-linear parameter sets NL3~\protect\cite{LKR.97} and
NL1~\protect\cite{RRM.86}.}
\begin {tabular}{ccccc}
                       &   DD-ME1   &  TW-99      & NL3     &   NL1 \\
\hline
\hline
$\rho_{\rm sat}$ (fm$^{-3}$) &   0.152    &   0.153     &   0.149 &   0.153 \\
E/A (MeV)            & -16.20     &  -16.25     &  -16.25 & -16.42 \\
K$_{0}$ (MeV)         & 244.5      & 240.0       & 271.8   & 211.3 \\
m$^{*}$               &   0.578    &   0.556     &   0.60  &   0.57 \\
a$_{4}$ (MeV)         &  33.1      &  32.5       &  37.9   &  43.7 \\
p$_{0}$ (MeV/fm$^{3}$)   &   3.26     &   3.22      &   5.92  &   7.0 \\
$\Delta$K$_{0}$ (MeV)   &-128.5      &-126.5       &  52.1   &  67.3
\end{tabular}
\label{tab2}
\end{table}
\begin {table}[!]
\begin {center}
\caption {Binding energies, charge radii and differences between neutron
and proton radii used to adjust the parameters of the DD-ME1 interaction.
The calculated values are compared with experimental data (in parentheses).}
\begin {tabular}{ccccccc}
          & & {E/A (MeV)} & & {$r_{ch}$ (fm)} & &{$r_n - r_p$ (fm)}\\ \hline
\hline
{$^{16}$O}   & & {-7.974 (-7.976)} & & {2.730 (2.730)} & &{-0.03       } \\
{$^{40}$Ca}  & & {-8.576 (-8.551)} & &{3.464 (3.485)} & &{-0.05       } \\
{$^{48}$Ca}  & & {-8.631 (-8.667)} & &{3.482 (3.484)} & &{ 0.19       } \\
{$^{90}$Zr}  & &{-8.704 (-8.710)} & &{4.294 (4.272)} & &{ 0.06       } \\
{$^{112}$Sn} & &{-8.501 (-8.514)} & &{4.586 (4.596)} & &{ 0.11       } \\
{$^{116}$Sn} & &{-8.516 (-8.523)} & &{4.616 (4.626)} & &{ 0.15 (0.12)} \\
{$^{124}$Sn} & &{-8.462 (-8.467)} & &{4.671 (4.674)} & &{ 0.25 (0.19)} \\
{$^{132}$Sn} & &{-8.352 (-8.355)} & &{4.720        } & &{ 0.27       } \\
{$^{204}$Pb} & &{-7.885 (-7.880)} & &{5.500 (5.486)} & &{ 0.18       } \\
{$^{208}$Pb} & &{-7.884 (-7.868)} & &{5.518 (5.505)} & &{ 0.20 (0.20)} \\
{$^{214}$Pb} & &{-7.764 (-7.772)} & &{5.568 (5.562)} & &{ 0.27       } \\
{$^{210}$Po} & &{-7.857 (-7.834)} & &{5.553        } & &{ 0.18       }
\end {tabular}
\label{tab3}
\end{center}
\end{table}
\begin {table}[!]
\begin {center}
\caption {Energy separation (in MeV) between spin-orbit partner states
in doubly closed-shell nuclei, calculated with the DD-ME1 and NL3
interactions, and compared with experimental data~\protect\cite{SO}.}

\begin {tabular}{cccccc}
    & &{ DD-ME1}& { NL3}& &{Exp.}   \\ \hline
\hline
$^{16}$O    & $\nu 1p$   &  6.316 &  6.482  & & 6.18    \\
            & $\pi 1p$   &  6.249 &  6.404  & & 6.32    \\ 
\hline
$^{40}$Ca   & $\nu 1d$   &  6.567 &  6.716  & & 6.00    \\
            & $\pi 1d$   &  6.507 &  6.630  & & 6.00    \\ 
\hline
$^{48}$Ca   & $\nu 1f$   &  7.689 &  7.542  & & 8.38    \\
            & $\nu 2d$   &  1.723 &  0.888  & & 2.02   \\ 
\hline
$^{132}$Sn  & $\nu 2d$   &  1.883 &  1.573  & & 1.65  \\
            & $\pi 1g$   &  6.244 &  6.230  & & 6.08  \\
        & $\pi 2d$   &  1.822 &  1.584  & & 1.75  \\ 
\hline
$^{208}$Pb  & $\nu 2f$   &  2.197 &  1.860  & & 1.77  \\
            & $\nu 1i$   &  6.839 &  6.813  & & 5.84  \\
        & $\nu 3p$   &  0.878 &  0.802  & & 0.90 \\
        & $\pi 2d$   &  1.647 &  1.525  & & 1.33 \\
        & $\pi 1h$   &  5.837 &  5.809  & & 5.56

\end {tabular}
\label{tab4}
\end{center}
\end{table}
\newpage
\end{document}